\documentclass{pinchcr}   
\usepackage{makeidx}
\usepackage{amsmath}
\usepackage{amssymb}
\usepackage{subfigure}
\usepackage{sidecap}

\newcommand{\pap}{\psi_{\rm app}}

\title{\normalsize{\emph{Fluctuating Nonlinear Oscillators}, Ed.~Mark
    I.~Dykman\\ (Oxford University Press, Oxford, 2012), Chapter 11:} \\\vskip30pt
\huge{\bf Collective Dynamics in Arrays of Coupled Nonlinear Resonators}}

\author{Ron Lifshitz}
\affiliation{Raymond and Beverly Sackler School of Physics \&
  Astronomy, Tel Aviv University, Tel Aviv 69978, Israel}
\author{Eyal Kenig and M.C. Cross}
\affiliation{Condensed Matter Physics 149-33, California Institute of
  Technology, Pasadena, CA 91125, USA}
\authors{2}

\makeindex
\begin{document}

\renewcommand{\eqref}[1]{eqn~(\ref{#1})}
\newcommand{\secref}[1]{Sec.~\ref{#1}}

\maketitle

\tableofcontents

\maintext

\chapter[Collective Dynamics in Arrays of Coupled Nonlinear
  Resonators]{Collective Dynamics in Arrays of Coupled Nonlinear
  Resonators\\
{\normalsize Ron Lifshitz, Eyal Kenig, and M.C. Cross}}

\section{Arrays of Nonlinear MEMS \& NEMS Resonators}
\label{sec:intro}

The study of collective nonlinear dynamics of coupled mechanical
resonators is regaining attention in recent years thanks to rapid
developments in the fields of microelectromechanical and
nanoelectromechanical systems (MEMS and NEMS)
\shortcite{RoukesPlenty,Cleland03}.  MEMS \& NEMS resonators are
typically characterized by very high frequencies, extremely small
masses, and weak damping. As such, they are naturally being developed
for a variety of applications such as sensing with unprecedented
accuracy~\shortcite{Rugar04,Ilic04,Ekinci04,Yang06,Li07,Naik09,Lee10}. NEMS,
in particular, are being developed also for studying fundamental
physics at small scales---exploring mesoscopic
phenomena~\shortcite{Schwab00,Weig04}, and even approaching quantum
behavior~\shortcite{LaHaye,Naik06,Rocheleau10,OConnell10}.  MEMS \&
NEMS resonators often exhibit nonlinear behavior in their dynamics, as
recently reviewed by Lifshitz and
Cross~\citeyear{LCreview08,LCreview10} and by
\shortciteN{Rhoads10}. This includes nonlinear resonant response
showing frequency pulling, multistability, and
hysteresis~\shortcite{C00,turner98,zaitsev,aldridge05,kozinsky07}, the
appearance of chaotic
dynamics~\shortcite{scheible02,Demartini07,Karabalin09,Kenig11}, as
well as the formation of extended~\shortcite{BR} and
localized~\shortcite{sato06,sato07,sato08} collective states in arrays
of coupled nonlinear resonators. Nonlinearities may be a nuisance in
actual applications, and schemes are being developed to avoid them, as
demonstrated, for example, by Kacem \emph{et
  al.}~\citeyear{Kacem09,Kacem10}. On the other hand, one can also
benefit from the existence of nonlinearity, for example in
mass-sensing applications~\shortcite{zhang02,Buks06}, in suppressing
noise induced phase diffusion, as suggested
by~\shortciteN{greywall94}, and in achieving self-synchronization of
large arrays, as proposed by Cross \emph{et
  al.}~\citeyear{sync1,sync2}. Nonlinearity is even proposed by Katz
\emph{et al.}~\citeyear{katz07,katz08} as a way to detect quantum
behavior in large mechanical systems.

Current technology enables the fabrication of large arrays, composed
of hundreds to tens of thousands of MEMS and NEMS devices, coupled by
electric, magnetic, or elastic forces. These arrays offer new
possibilities for quantitative studies of nonlinear dynamics in
systems with an intermediate number of degrees of freedom---much
larger than one can deal with in macroscopic experiments, yet much
smaller than one confronts when considering nonlinear aspects of
phonon dynamics in a crystal. Our studies of collective nonlinear
dynamics of MEMS and NEMS were originally motivated by the experiment
of \shortciteN{BR}. These studies have led to a quantitative
understanding of the collective response of arrays of nonlinear
resonators, providing explicit bifurcation diagrams that explain the
transitions between different extended modes of an array as the
strength and frequency of the external drive are varied
quasistatically~\shortcite{LC,BCL}. We have considered more general
issues such as the nonlinear competition between extended modes, or
patterns, of the system---when many such patterns are simultaneously
stable---as the external driving parameters are changed abruptly or
ramped as a function of time~\shortcite{kenig09}. We have also studied
the formation, stability, and rich dynamics of intrinsically localized
modes \shortcite{kenigILM}. Furthermore, we have investigated the
synchronization that may occur in coupled arrays of
\emph{non-identical} nonlinear oscillators, based on the ability of
nonlinear oscillators to tune their frequency by changing their
oscillation amplitude~\shortcite{sync1,sync2}. 

The purpose of this chapter is to provide a review of the collective
dynamical phenomena observed in these different systems, while
highlighting the common concepts and theoretical tools that we have
developed for dealing with them. We assume that the reader is familiar
with the basic dynamical phenomena associated with single nonlinear
resonators. The unfamiliar reader is encouraged to consult our
previous review on the subject \shortcite{LCreview08}, or its revised
version \shortcite{LCreview10}. In \secref{sec:eom} we describe the
equations of motion that are used to model arrays of nonlinear MEMS
and NEMS resonators, for different experimental realizations. We then
give two examples of the derivation and then application of
\emph{discrete amplitude equations} for treating arrays of
resonators---in \secref{sec:LC} we study the resonant nonlinear
response of arrays to parametric excitation, and in \secref{sec:sync}
we discuss the question of synchronization. We conclude with two
examples of the derivation and then application of \emph{continuous
  amplitude equations} for treating large arrays of resonators---in
\secref{sec:patterns} for investigating pattern selection, and in
\secref{sec:ILM} for the study of intrinsically localized modes.  In
place of a formal concluding section, we wish to emphasize at the
outset that all the results obtained from analyzing the different
amplitude equations are in excellent agreement with numerical
solutions of the underlying equations of motion. This upholds the
validity of using such reduced descriptions for complex systems, whose
original description is given in terms of coupled nonlinear ordinary
differential equations. Furthermore, our numerical simulations of the
equations of motion suggest that the predicted effects can be observed
in arrays of real MEMS and NEMS resonators, thus motivating new
experiments in these systems.

\section{Equations of Motion and Basic Assumptions}
\label{sec:eom}

Typical MEMS and NEMS resonators are characterized by extremely high
fre\-quen\-cies---now going beyond 10
GHz~\shortcite{HZMR03,cleland04,Weinstein10}---and relatively weak
dissipation, with quality factors $Q$ in the range of $10^{2}-10^{5}$.
For such devices, under external driving conditions, transients die
out rapidly, making it is easy to acquire sufficient data to
characterize the steady-state well.  This, and the fact that weak
dissipation and weak nonlinearity can be treated perturbatively, are a
great advantage for quantitative comparison between theory and
experiment.

\subsection{Modeling a single nonlinear resonator}

A typical single resonator is described after appropriate scaling by a
dimensionless equation of motion of the form~\shortcite{LCreview08}
\index{equation of motion!single resonator}
\begin{equation}
  \label{eq:parascaledduffing}
  \ddot x + Q^{-1}\dot x + \left[1+ H \cos\omega_P t\right] x + x^3 +
  \eta x^2\dot x = G\cos\left(\omega_D t + \phi_g\right).
\end{equation}
We typically use the fact that damping $Q^{-1}$\index{damping!linear}
is much smaller than the resonant frequency, which has been scaled
here to 1, to define the small expansion parameter $\epsilon=Q^{-1}$.
The term proportional to $H$ on the left hand side is an external
drive that modulates the spring constant. This is a parametric
drive---a term that is proportional to the displacement $x$ as well as
to the strength of the drive.\index{parametric drive} The term
proportional to $G$ on the right-hand side is the standard direct
drive, possibly shifted by a phase $\phi_g$ with respect to the
parametric drive. The coefficient of the nonlinear $x^3$
Duffing\index{Duffing nonlinearity} term has been scaled to 1, and a
nonlinear damping term~[Dykman and
Krivoglaz~\citeyear{Dykman75,Dykman84}] with coefficient $\eta$ is
also added.\index{damping!nonlinear} For parametric drive, we normally
consider the largest excitation effect that occurs when the pump
frequency $\omega_P$ is close to twice the resonant frequency of the
resonator.  We therefore take $\omega_p=2+\epsilon\Omega_{P}$, and
take the drive amplitude to scale as the damping by setting
$H=\epsilon h$. The amplitude of the direct drive is scaled as
$G=\epsilon^{3/2} g$, and its frequency is set an amount
$\epsilon\Omega_{D}$ away from the resonant frequency.

The scaled equation of motion that we then obtain is of the form
\index{equation of motion!single resonator!scaled}
\begin{equation}
  \label{eq:paraDuffing}
  \ddot{x} + \epsilon\dot{x} + \left(1 + \epsilon
    h\cos\left[\left(2+\epsilon\Omega_{P}\right)t\right]\right)x +
  x^{3} + \eta x^{2}\dot{x} = \epsilon^{3/2} |g| \cos\left[\left(1 +
      \epsilon\Omega_D\right) t
    + \phi_g\right],
\end{equation}
where we use $g = |g|e^{i\phi_g}$ to denote a complex drive
amplitude. Ignoring transients, which as explained above decay very
rapidly, the solutions to such equations of motion are of the form
$x=\sqrt\epsilon~\Re\{A(T)e^{it}\}$ plus corrections of higher order
in $\epsilon$, where\index{secular perturbation theory} secular
perturbation theory is used to yield the equation that governs the
slow dynamics of the complex amplitude $A(T)$. The variable
$T=\epsilon t$ is the slow time scale upon which the interesting
nonlinear dynamics takes place. Please refer to Lifshitz and
Cross~\citeyear{LCreview08,LCreview10} for more details and for many
examples of the use of this approach. We only wish to remind the
reader that additional nonlinear terms, up to third order in $x$ or
$\dot x$ such as $x^2$ and $x\dot x^2$, that seem to be missing in
\eqref{eq:paraDuffing}, merely conspire to renormalize the effective
parameters in the slow equation for $A(T)$, but do not affect the
actual form of this equation. We therefore ignore all such terms as
they have no effect on the actual nature of the solutions that we
study.

\subsection{Modeling an array of nonlinear resonators}
\label{intro-array}

\citeN{LC} had originally modeled a 1-dimensional array of
parametrically driven coupled nonlinear resonators, motivated by the
particular experiment of \shortciteN{BR}, in which an array of 67
doubly-clamped micromechanical gold beams was parametrically excited
by modulating the strength of an externally-controlled electrostatic
coupling\index{coupling!electrostatic} between neighboring beams. We
used a set of coupled equations of motion of the form
\index{equation of motion!one-dimensional array}
\begin{eqnarray}\label{LCeom}
\ddot{u}_{n} & + u_{n} - \tfrac{1}{2} Q^{-1}(\dot{u}_{n+1} - 2\dot{u}_{n} +
\dot{u}_{n-1}) + \tfrac{1}{2}\left(D + H\cos\omega_p t\right)(u_{n+1}
-2u_{n}+u_{n-1})\nonumber\\
& + u_{n}^{3} -\tfrac{1}{2}\eta\bigl[(u_{n+1} -
u_{n})^{2}(\dot{u}_{n+1} - \dot{u}_{n}) - (u_{n} - u_{n-1})^{2}
(\dot{u}_{n} - \dot{u}_{n-1})\bigr]=0,
\end{eqnarray}
where $u_n(t)$ describes the deviation of the $n^{th}$ resonator from
its equilibrium, with $n=1\ldots N$, and fixed boundary conditions
$u_{0}=u_{N+1}=0$.\index{boundary conditions!fixed} Detailed arguments
for the choice of terms introduced into these particular equations of
motion are discussed by~\shortciteN{LC}. We only note that they
contain nearest-neighbor \emph{linear coupling} which is both
\emph{reactive},\index{coupling!linear reactive} proportional to the
relative displacements, and \emph{dissipative},%
\index{coupling!linear dissipative} proportional to the relative
velocities; as well as \emph{nonlinear dissipative coupling},%
\index{coupling!nonlinear dissipative} proportional to the square of
the relative displacements and to the relative velocities.

A simpler model, suitable in many other situations, is to take the
equation of motion of each resonator to be as
in~(\ref{eq:parascaledduffing}) with the addition of only a linear
reactive coupling\index{coupling!linear reactive} term to its two
neighbors.
The equations of motion then take the form
\index{equation of motion!one-dimensional array}
\begin{equation}
  \ddot{u}_{n} + Q^{-1}\dot{u}_{n} + (1 + H\cos\omega_{P}t)
  u_{n} + u_{n}^{3}  + \eta u_{n}^{2}\dot{u}_{n} + \tfrac{1}{2}  D(u_{n+1}-2u_{n}+u_{n-1}) =0,
\label{eq:eom}
\end{equation}
where in both cases one could add the direct drive, proportional to
$G$, that was considered earlier in \eqref{eq:parascaledduffing}.

Finally, to model an array of oscillators---having a
frequency-independent source of energy that sustains their
oscillations---rather than simple resonators that respond resonantly
to an external frequency-dependent drive, we consider a slight
modification of \eqref{eq:eom}, given by
\index{equation of motion!one-dimensional array}
\begin{equation}
\ddot{u}_{n} + \omega_{n}^{2}u_{n} - \nu(1-u_{n}^{2})\dot{u}_{n} +
au_{n}^{3} + \tfrac{1}{2}  D(u_{n+1}-2u_{n}+u_{n-1})
= 0. \label{eq:syncmodel}
\end{equation}
In this case both the parametric drive and the direct drive are
omitted.  Instead, we introduce a \emph{negative} linear
damping\index{damping!linear negative} with coefficient $\nu$, which
represents an energy source to sustain the oscillations, while keeping
the positive nonlinear damping\index{damping!nonlinear} so that the
oscillation amplitude saturates\index{nonlinear saturation} at a
finite value. We use a different scaling than before to set this
saturation value to be of order unity, and therefore must reintroduce
an explicit coefficient $a$ in front of the $x^3$ Duffing
term,\index{Duffing nonlinearity} which can no longer be scaled to
unity.  One can implement such an effect with an electronic feedback
loop, sensing each oscillator velocity and driving the oscillator with
an appropriate phase~\shortcite{Feng08}. The first three terms of
\eqref{eq:syncmodel} comprise a so-called \emph{van der Pohl}
oscillator. \index{van der Pohl oscillator} Note that in anticipation
of our study of synchronization of coupled oscillators in
\secref{sec:sync} below, we have assumed that the uncoupled
oscillators can generally have non-identical linear frequencies
$\omega_n$.

The equations of motion for particular experimental implementations
might have different terms, although we expect all will have positive
or negative Duffing terms;\index{Duffing nonlinearity} linear and
possibly also nonlinear damping;\index{damping!nonlinear}%
\index{damping!linear} linear and possibly also nonlinear coupling,
\index{coupling!nonlinear reactive}%
\index{coupling!nonlinear dissipative}%
\index{coupling!linear reactive}%
\index{coupling!linear dissipative}%
which may be either reactive or dissipative; and some source of energy
to sustain the oscillations. In many cases, although not always, once
we transform to the reduced description describing the slow
modulation of the modes (see below), the differences between these
different models will not lead to qualitatively new effects.

\section{Discrete Amplitude Equations:\\
Example I -- Collective response to parametric excitation}
\label{sec:LC}
\subsection{Deriving the equations}

As in the case of
a single resonator in \eqref{eq:paraDuffing}, we suppose $Q$ is large
and take $\epsilon=Q^{-1}$ as a small expansion parameter. Again, we
take $H=\epsilon h$, and in addition also take $D=\epsilon d$ so that
the width of the frequency band of normal modes is also small. This is
not quite how \shortciteN{LC} treated the coupling, but it is simpler yet
equivalent up to the order of the expansion in $\epsilon$ that we
require.  The equations of motion~(\ref{eq:eom}) then become
\index{equation of motion!one-dimensional array!scaled}
\begin{equation}
  \label{eq:neweom}
  \ddot{u}_{n} + \epsilon\dot{u}_{n} + \left(1 + \epsilon
  h\cos\left[\left(2 + \epsilon\Omega_{P}\right)t\right]\right)u_{n} 
  + \tfrac{1}{2} \epsilon d(u_{n+1}-2u_{n}+u_{n-1}) + u_{n}^{3} 
  + \eta u_{n}^{2}\dot{u}_{n}=0.
\end{equation}
We expand $u_{n}(t)$ as a sum of standing wave modes with slowly
varying amplitudes. The nature of the standing wave modes will depend
on the conditions at the ends of the array of resonators. In the
experiment of \shortciteN{BR} there where $N$ mobile beams with a
number of identical immobilized beams at each end. These
conditions\index{boundary conditions!fixed} can be implemented in a nearest
neighbor model by taking two additional resonators, $u_{0}$ and
$u_{N+1}$ and assuming
\begin{equation}
u_{0}=u_{N+1}=0. \label{eq:pinnedbc}%
\end{equation}
The standing wave modes are then%
\begin{equation}
u_{n}=\sin(nq_{m})\qquad\text{with}\qquad
q_{m}=\frac{m\pi}{N+1},\ m=1\ldots N.
\label{eq:sinsw}
\end{equation}
On the other hand, for an array of $N$ resonators with free ends there
is no force from outside the array. For the nearest neighbor model this
can be imposed again by taking two additional resonators, but now
with the conditions\index{boundary conditions!free}
\begin{equation}
u_{0}=u_{1};\qquad u_{N}=u_{N+1}. \label{eq:freebc}
\end{equation}
The standing wave modes are now%
\begin{equation}
u_{n} = \cos\left[\left(n-\tfrac{1}{2}\right)q_{m}\right]\qquad \text{with}\qquad q_{m} =
\frac{m\pi}{N},\ m=0\ldots N-1. \label{eq:cossw}
\end{equation}
For our illustration we will take eqns~(\ref{eq:pinnedbc},
\ref{eq:sinsw}).

To treat the equations of motion~(\ref{eq:neweom}) analytically, we
use secular perturbation theory combined with a multiple scales
analysis,\index{multiple scales} taking advantage of the natural
separation of time scales that occurs in our physical system---the
fast oscillations of the resonators at half the drive frequency are
characterized by the fast time variable $t$, whereas the slow
variation of the amplitudes of these oscillations is associated with
transient times, characterized by the damping rate $Q^{-1}$, or
$\epsilon$, giving rise to a well-separated slow time variable
$T=\epsilon t$. This approach is used throughout this review, and was
described in great detail in our previous
review~\cite{LCreview08,LCreview10}. Thus, we introduce the
ansatz\index{multiple scales!ansatz}
\begin{equation}
u_{n}(t)=\epsilon^{1/2}\frac{1}{2}\sum_{m=1}^{N}\left(
A_{m}(T)\sin (nq_{m})e^{it}+c.c.\right)
+\epsilon^{3/2}u_{n}^{(1)}(t)+\ldots,\
n=1\ldots N, \label{eq:ansatz}
\end{equation}
where $c.c.$ stands for the complex conjugate. The lowest order
contribution to this solution is based on the normal mode
solutions~(\ref{eq:sinsw}) of the linear equations of motion, allowing
the complex mode amplitudes $A_m(T )$ to vary slowly in time (as in a
rotating frame in the complex plane),\index{rotating frame} due to the
effect of all the other terms in the equation. As we shall immediately
see, the slow temporal variation of $A_m(T )$ also allows us to ensure
that the perturbative correction $u_n^{(1)}(t)$, as well as all
higher-order corrections to the solution~(\ref{eq:ansatz}), do not
diverge as they do if one uses naive perturbation theory.

Using the relation
\begin{equation}
\dot{A}_{n}={\frac{dA_{n}}{dt}}=\epsilon{\frac{dA_{n}}{dT}}\equiv\epsilon
A_{n}^{\prime}, \label{eq:adot}%
\end{equation}
and denoting a time derivative with respect to the slow time $T$ by a
prime, we substitute the trial solution~(\ref{eq:ansatz}) into the
equations of motion (\ref{eq:neweom}) term by term. Up to order
$\epsilon^{3/2}$ we have,
\begin{subequations}
\begin{align}
&\ddot{u}_{n}  
=\epsilon^{1/2}\frac{1}{2}\sum_{m}\sin(nq_{m})\left(
[-A_{m}+2i\epsilon A_{m}^{\prime}]e^{it}+c.c.\right)
+\epsilon^{3/2}\ddot
{u}_{n}^{(1)}(t),\label{eq:xnddot}\\
&\epsilon\dot{u}_{n}  =\epsilon^{3/2}\frac{1}{2}\sum_{m}\sin
(nq_{m})\left(  iA_{m}e^{it}+c.c.\right), \label{eq:xndot}\\%
&\epsilon \frac{d}{2} (u_{n+1}-2u_{n}+u_{n-1})=-\epsilon^{3/2}
\frac{d}2\sum_{m}2\sin^{2}\left(  \frac{q_{m}}{2}\right)
\sin(nq_{m})\left( A_{m}e^{it}+c.c.\right),\\
\label{eq:cube}\nonumber
&u_{n}^{3}  =\epsilon^{3/2}\frac{1}{8}\sum_{j,k,l}\sin(nq_{j})\sin
(nq_{k})\sin(nq_{l})\left(  A_{j}e^{it}+c.c.\right)  \left(
A_{k}e^{it}+c.c.\right)
\left(  A_{l}e^{it}+c.c.\right) \nonumber\\
&\hphantom{u_{n}^{3}}  =\epsilon^{3/2}\frac{1}{32}\sum_{j,k,l}\left\{  \sin[n(-q_{j}+q_{k}%
+q_{l})]+\sin[n(q_{j}-q_{k}+q_{l})] +\sin[n(q_{j}+q_{k}-q_{l})]\right. \nonumber\\
&  \hphantom{=\epsilon^{3/2}\frac{1}{32}\sum_{j,k,l}} \left.
-\sin[n(q_{j}+q_{k}+q_{l})]\right\}
\left\{  A_{j}A_{k}A_{l}e^{3it}+3A_{j}A_{k}A_{l}^{\ast}%
e^{it}+c.c.\right\},
\end{align}
and
\begin{align}
\eta u_{n}{}^{2}\dot{u}_{n}  
&=\epsilon^{3/2} \frac{\eta}{32} \sum _{j,k,l}\left\{
\sin[n(-q_{j}+q_{k}+q_{l})]+\sin[n(q_{j}-q_{k}+q_{l})] +\sin[n(q_{j}+q_{k}-q_{l})]\right.
\nonumber\label{eq:dotcube}\\
&\left. -\sin[n(q_{j}+q_{k}+q_{l})] \right\} \left(  A_{j}e^{it}+c.c.\right)  \left(
A_{k}e^{it}+c.c.\right) \left(  iA_{l}e^{it}+c.c.\right).
\end{align}

The order $\epsilon^{1/2}$ terms cancel, and at order
$\epsilon^{3/2}$ we get $N$ equations of the form
\end{subequations}
\begin{equation}
\ddot{u}_{n}^{(1)}+u_{n}^{(1)}=\sum_{m}\left(  m^{th}\ \mathrm{secular\ term}%
\right)  e^{it}+\mathrm{other\ terms},\label{eq:order3/2}%
\end{equation}
where the left-hand sides are uncoupled linear harmonic resonators,
with a frequency unity. On the right-hand sides we have $N$ secular
terms\index{secular term} which act to drive the resonators
$u_{n}^{(1)}$ at their resonance frequencies. As Lifshitz and Cross
\citeyear{LCreview08,LCreview10} did for all their single resonator
examples, here too we require that all the secular terms vanish so
that the $u_{n}^{(1)}$ remain finite. This is the necessary
\emph{solvability condition},\index{solvability condition} required to
obtain equations for the slowly varying amplitudes $A_{m}(T)$. To
extract the equation for the $m^{th}$ amplitude $A_{m}(T)$ we make use
of the orthogonality of the modes, multiplying all the terms by
$\sin(nq_{m})$ and summing over $n$.  We find that the coefficient of
the $m^{th}$ secular term, which is required to vanish, is given by
\index{amplitude equation!discrete}
\begin{equation}\label{eq:Am1}
  \boxed{
    -2i{\frac{dA_{m}}{{dT}}-iA_{m}+} 2d\sin^{2}\left( \frac{q_{m}}
      {2}\right)  A_{m}-\frac{1}{2}hA_{m}^{\ast}e^{i\Omega_{P}T}
    -{\frac{3+i\eta}{16}}\sum_{j,k,l}A_{j}A_{k}A_{l}^{\ast}\Delta_{jkl;m}^{(1)}=0
  },
\end{equation}
where we have used the $\Delta$ function introduced by \shortciteN{LC}, defined in
terms of Kronecker deltas as
\begin{equation}\label{eq:Delta1}
\begin{split}
\Delta_{jkl;m}^{(1)}&=\delta_{-j+k+l,m}-\delta_{-j+k+l,-m}-\delta
_{-j+k+l,2(N+1)-m}\\
&+\delta_{j-k+l,m}-\delta_{j-k+l,-m}-\delta_{j-k+l,2(N+1)-m}\\
&+\delta_{j+k-l,m}-\delta_{j+k-l,-m}-\delta_{j+k-l,2(N+1)-m}\\
&-\delta_{j+k+l,m}+\delta_{j+k+l,2(N+1)-m}-\delta_{j+k+l,2(N+1)+m},
\end{split}
\end{equation}
and have exploited the fact that it is invariant under any permutation
of the indices $j$, $k$, and $l$.  The $\Delta$ function ensures the
conservation of lattice momentum---the conservation of momentum to
within the non-uniqueness of the specification of the normal modes due
to the fact that $\sin(nq_{m})=\sin(nq_{2k(N+1)\pm m})$ for any
integer $k$. The first Kronecker delta in each line is a condition of
direct momentum conservation, and the other two are the so-called
umklapp conditions\index{umklapp condition} where only lattice
momentum is conserved.

As for the single resonator~\cite{LCreview08}, we again try a
steady-state solution,\index{steady-state solution!scaled} this time of the
form
\begin{equation}
A_{m}(T)=a_{m}e^{i\left(  {\frac{\Omega_{P}}{2}}\right)  T},\label{eq:Am}
\end{equation}
so that the solutions to the equations of motion (\ref{eq:neweom}),
after substitution of (\ref{eq:Am}) into (\ref{eq:ansatz}), become
\index{steady-state solution}
\begin{equation}
u_{n}(t)=\epsilon^{1/2}\frac{1}{2}\sum_{m}\left(
a_{m}\sin(nq_{m}) e^{i\left(1 + {\frac{\epsilon\Omega_{P}}{2}}\right)
  t}+c.c.\right) + O(\epsilon^{3/2}),\label{eq:solj}
\end{equation}
where all modes are oscillating at half the parametric excitation
frequency, $\omega_P=2+\epsilon\Omega_P$.

Substituting the steady state solution (\ref{eq:Am}) into the
equations (\ref{eq:Am1}) for the time-varying amplitudes $A_{m}(T)$,
we obtain the equations for the \emph{time-independent} complex amplitudes
$a_{m}$\index{amplitude equation!discrete!time-independent}
\begin{equation}
  \label{eq:ampeq}
  \boxed{
    \left[\Omega_{P} + 2d\sin^{2}\left(
        \frac{q_{m}}{2}\right)-i\right]  a_{m}-\frac{h}{2}a_{m}^{\ast}
    -\frac{3+i\eta}{16}\sum_{j,k,l}a_{j}a_{k}a_{l}^{\ast}
    \Delta_{jkl;m}^{(1)}=0
  }.
\end{equation}

Note that the first two terms on the left-hand side indicate that the
linear resonance frequency is not obtained for $\Omega_P=0$, but
rather for $\Omega_P + 2d\sin^{2}\left(q_m/2\right)=0$. In terms of
the unscaled parameters, this implies that the resonance frequency of
the $m^{th}$ mode is $\omega_m = 1 - D\sin^{2}\left(q_m/2\right)$,
which to within a correction of $O(\epsilon^2)$ is the same as the
expected dispersion relation\index{dispersion relation}
\begin{equation}
  \label{eq:arraydispersion}
  \omega_m^2 = 1 - 2D \sin^{2}\left(\frac{q_m}{2}\right).
\end{equation}

Equations~(\ref{eq:Am1}) and (\ref{eq:ampeq}) are the main result of
the calculation. We have managed to replace $N$ coupled differential
equations (\ref{eq:eom}) for the resonator coordinates $u_{n}(t)$ by
$N$ coupled differential equations~(\ref{eq:Am1}) for the slowly
varying mode amplitudes $A_m(T)$, and then by $N$ coupled algebraic
equations (\ref{eq:ampeq}) for the time-independent mode amplitudes
$a_{m}$. All that remains, in order to obtain the overall collective
response of the array as a function of the parameters of the original
equations of motion (\ref{eq:eom}), is to solve these coupled algebraic equations.

\subsection{Analyzing and solving the equations}

A number of simple results can immediately be stated. First, one can
easily verify that for a single resonator ($N=j=k=l=m=1$), the general
equation (\ref{eq:ampeq}) reduces to the single-resonator equation
treated by Lifshitz and Cross \citeyear{LCreview08,LCreview10}, as
$\Delta_{111;1}=4$. Next, one can also see that the trivial solution,
$a_{m}=0$ for all $m$, always satisfies the equations, though, as
Lifshitz and Cross \citeyear{LCreview08,LCreview10} showed in the case
of a single resonator, it is not always a stable solution. Finally,
one can also verify that whenever for a given $m$,
$\Delta_{mmm;j}^{(1)}=0$ for all $j\neq m$, then a single-mode
solution exists with $a_{m}\neq0$ and $a_{j}=0$ for all $j\neq m$.
These single-mode solutions have the same type of elliptical shape of
the single-resonator solution. Note that generically
$\Delta_{mmm;m}^{(1)}=3$, except when umklapp conditions are
satisfied.

\begin{figure}[b]
\subfigure[]{\includegraphics[width=0.32\textwidth]{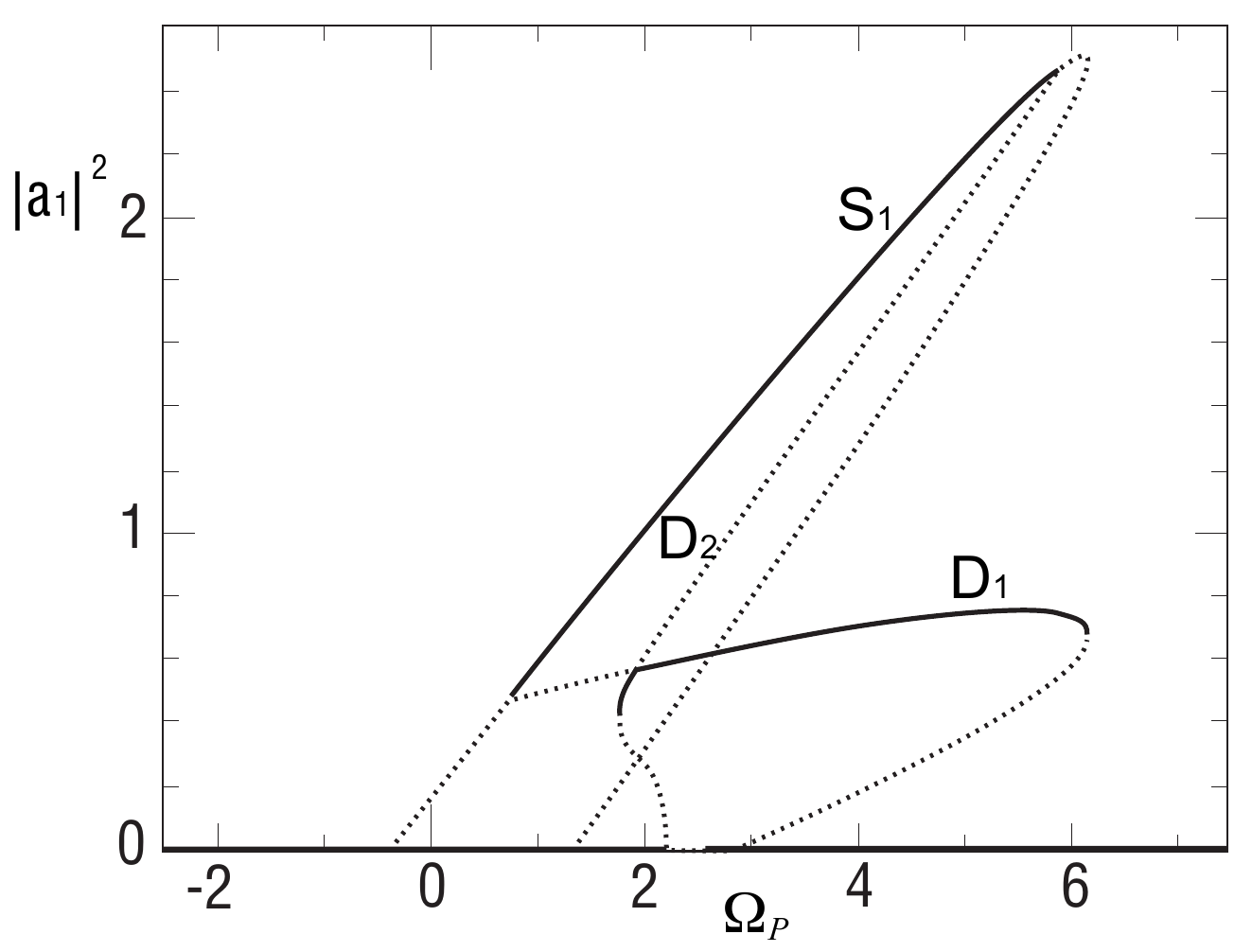}}%
\subfigure[]{\includegraphics[width=0.32\textwidth]{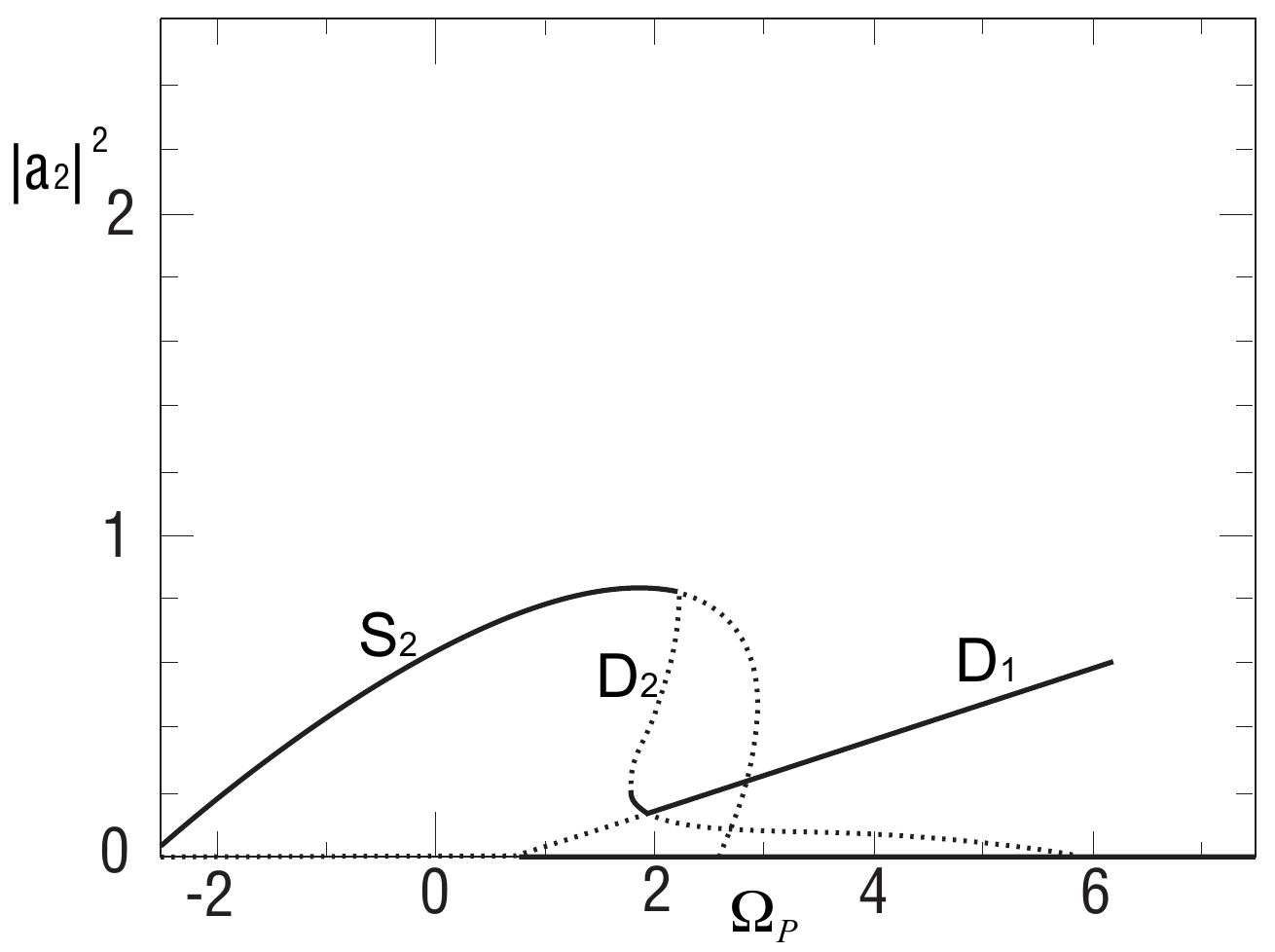}}%
\hskip10pt
\subfigure[]{\includegraphics[width=0.29\textwidth]{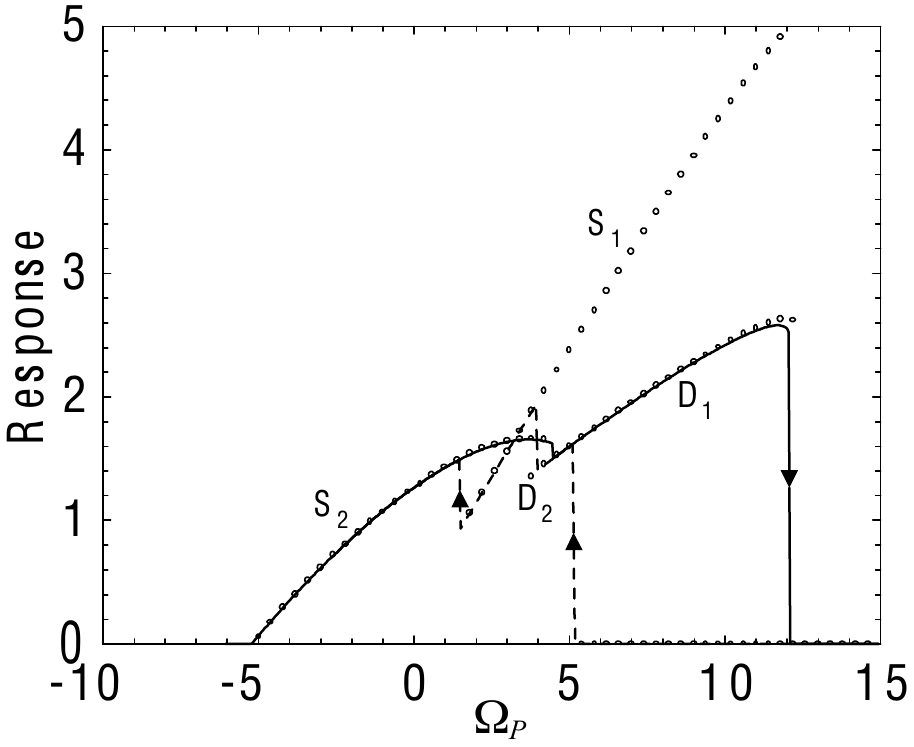}}%
\caption{Response intensity of two resonators as a function of
  frequency $\Omega_P$, for a particular choice of the equation
  parameters. (a) shows $|a_{1}|^{2}$, and (b) shows $|a_{2}|^{2}$, 
  with solid curves indicating stable solutions and
  dashed curves indicating unstable solutions. 
  (c) Comparison of stable solutions, obtained algebraically (small
  circles), with a numerical integration of the equations of motion
  (\ref{eq:eom}) (solid curve - frequency swept up; dashed curve -
  frequency swept down) showing\index{hysteresis} hysteresis in 
  the response. Plotted is the averaged response intensity, defined in the text.
  In all figures, the two elliptical
  single-mode solution branches are labeled $S_{1}$ and $S_{2}$, and the
  two double-mode solution branches are labeled $D_{1}$ and $D_{2}$. 
  From Lifshitz and Cross (2003). Copyright (2003) American Physical Society.
  \label{2beams}}
\end{figure}

Additional solutions, involving more than a single mode, exist in
general but are hard to obtain analytically. \shortciteN{LC}
calculated these multi-mode solutions explicitly for the case of two
and three resonators, for the model they considered, by finding the
roots of the coupled algebraic equations numerically. We present some
of their results to illustrate the type of behavior that occurs,
although the precise details will be slightly different in the model
used here. In Fig.~\ref{2beams} we show the solutions for the response
intensity of two resonators as a function of frequency, for a
particular choice of the equation parameters. Figure~\ref{2beams}(a)
shows the square of the amplitude of the symmetric mode $a_{1}$,
whereas Fig.~\ref{2beams}(b) shows the square of the amplitude of the
antisymmetric mode $a_{2}$. Solid curves indicate stable solutions and
dashed curves indicate unstable solutions. Two elliptical single-mode
solution branches, similar to the response of a single resonator are
easily spotted. These branches are labeled by $S_{1}$ and
$S_{2}$. \shortciteN{LC} give the analytical expressions for these two
solution branches. In addition, there are two double-mode solution
branches, labeled $D_{1}$ and $D_{2}$, involving the excitation of
both modes simultaneously.  Note that the two branches of double-mode
solutions intersect at a point where they switch their stability.

With two resonators there are regions in frequency where three stable
solutions can exist.\index{multistability} If all the stable solution
branches are accessible experimentally then the observed effects of
hysteresis\index{hysteresis} might be more complex than in the simple
case of a single resonator.  This is demonstrated in
Fig.~\ref{2beams}(c) where the algebraic solutions are compared with a
numerical integration of the differential equations of motion
(\ref{eq:eom}) for two resonators. The response intensity, plotted
here, is the time and space averages of the square of the resonator
displacements $(\langle u_{1}^{2}\rangle + \langle
u_{2}^{2}\rangle)/2$, where the angular brackets denote time
average. A solid curve shows the response intensity for an upward
quasistatic frequency sweep, and a dashed curve shows the response
intensity for a downward sweep.%
\index{quasistatic sweep!of drive frequency} Small circles show the
response intensity, as calculated for the stable regions of the four
\emph{algebraic} solution branches shown in Figs.~\ref{2beams}(a) and
(b), demonstrating the great utility of the slow amplitude
equations. With the analytical solution in the background, one can
easily understand all the discontinuous jumps, as well as the
hysteresis\index{hysteresis} effects, that are obtained in the
numerical solution of the equations of motion.  Note that the $S_{1}$
branch is missed in the upward frequency sweep and is only accessed by
the system in the downward sweep.  One could trace the whole stable
region of the $S_{1}$ branch by changing the sweep direction after
jumping onto the branch, thereby climbing all the way up to the end of
the $S_{1}$ branch.  These kinds of changes in the direction of the
quasistatic sweep whenever one jumps onto a new branch are essential
if one wants to trace out as much of the solution as
possible---whether in real experiments or in numerical simulations.

\subsection{Brief survey of applications}

Discrete amplitude equations like the ones derived
here~(\ref{eq:ampeq}) are useful mainly when studying small
arrays. Nevertheless, the insight gained by studying small arrays, of
even two or three resonators, provides better understanding of the
dynamics of large arrays, which can be studied directly by numerically
integrating the starting equations of motion~(\ref{eq:eom}). Indeed,
all the features of the original experiment of \citeN{BR} were
qualitatively reproduced by \citeN{LC} by numerically integrating
their original equations of motion~(\ref{LCeom}). But, it was their
analytical study of small arrays, that allowed them to provide an
explanation for the observed features: (1) The response of the system
at frequencies above the top edge of the band was attributed to the
positive frequency pulling\index{frequency pulling} coming from
the\index{Duffing nonlinearity} Duffing nonlinearity; (2) The fact
that only a few features are observed in a%
\index{quasistatic sweep!of drive frequency} monotonic quasistatic
frequency scan, rather than $N$ resonance peaks for the $N$ normal
modes, was explained by the fact that a solution branch is followed
quasistatically as long as it is stable, often skipping many other
solutions that are simultaneously stable,\index{multistability} as
demonstrated above with 2 resonators; and (3) The abrupt jumps in the
response were identified as stemming from bifurcation points where a
certain solution branch ends, as in the saddle-node
bifurcation\index{bifurcation!saddle-node} at the end of the $D_1$
branch in Fig.~\ref{2beams}, or simply loses its stability, as for the
two $S$ branches in Fig.~\ref{2beams}, in either case requiring the
system to switch abruptly to a different branch.

The curious reader is encouraged to consult additional articles, where
the methods presented in this section were used to study advanced
features in the dynamics of small numbers of coupled
resonators. \shortciteN{Karabalin09} used discrete amplitude equations
for the two normal modes of a pair of resonators, similar to
\eqref{eq:Am1}, to assist in their numerical modeling of period
doubling and a transition to chaos,\index{chaos} which they observed
experimentally. \shortciteN{Kenig11} used discrete amplitude equations
to identify homoclinic orbits\index{homoclinic orbits} in the slow
dynamics and assess the possibility of obtaining chaotic dynamics via
the Melnikov\index{Melnikov} approach. Finally, \shortciteN{BTA}
demonstrated the use of a pair of parametrically driven resonators as
a novel amplifier, whose operation is based on very sensitive control
of the bifurcation diagram of the response of two resonators, via an
input signal that is fed into the coupling $D$ between the
resonators. The Supplementary Material of \shortciteN{BTA} provides a
detailed analysis of the operation of this so-called
\emph{Bifurcation-Topology Amplifier},%
\index{bifurcation-topology amplifier} using a set of discrete
amplitude equations like the ones developed here.

\section{Discrete Amplitude Equations:\\
Example II -- Synchronization of nonlinear oscillators}
\label{sec:sync}
\index{synchronization|(}

\subsection{Deriving the equations}

Although synchronization is often put forward as an example of the
importance of understanding nonlinear phenomena, the intuition for it,
and indeed the subsequent mathematical discussion, often reduces to
simple linear ideas. For example, the famous example of Huygens's
clocks \shortcite{Bennett02}\index{Huygens clocks} can be understood
in terms of a linear coupling of the two pendulums through the common
mounting support. It is then the larger damping of the symmetric mode
(coming from the larger, dissipative motion of the common support)
compared with the antisymmetric mode that leads, at long times, to a
synchronized state of the two pendulums oscillating in antiphase.  The
nonlinearity in the system is simply present in the individual motion
of each pendulum; specifically in the mechanism to sustain the
oscillations. Without the drive, the oscillators would still become
synchronized through the faster decay of the even mode, albeit in a
slowly decaying state. Rather than this mode-dependent dissipation
mechanism, one might expect synchronization to arise from the
intrinsically nonlinear effect of the frequency pulling of one
oscillator by another. Furthermore, the model describing the two
Huygens pendulums, as well as most other models used to show
synchronization, has\index{coupling!linear dissipative} dissipative
coupling between the oscillators.  In contrast, many physical
situations have mainly reactive coupling. Consequently, Cross \emph{et
  al.}~\citeyear{sync1,sync2} proposed and analyzed a model for
synchronization, given by \eqref{eq:syncmodel}, involving
\emph{reactive coupling}\index{coupling!linear reactive} between the
oscillators, which then leads to synchronization through
\index{Duffing nonlinearity}\index{frequency pulling}\emph{nonlinear
  frequency pulling}.

We follow Cross \emph{et al.} \citeyear{sync1,sync2} and consider the
system of oscillators defined by \eqref{eq:syncmodel}, assuming that
the linear frequencies of the oscillators are distributed near unity
such that
\begin{equation}
\omega_{n}^{2 }= 1 + \Delta_{n}, \qquad\text{with}\qquad |\Delta_{n}|\ll1.
\end{equation}
This allows us to study the situation in which the equations of motion
are dominated by the terms describing simple harmonic oscillators at
frequency one, and the time dependence remains close to $e^{\pm
  it}$. The interesting dynamics should then be captured by a discrete
set of coupled amplitude equations for the deviations of the
individual oscillators from simple harmonic oscillation at frequency
1. To that end we assume that all corrections in \eqref{eq:syncmodel}
to a set of uncoupled harmonic oscillators of frequency 1 are small.
To formalize this smallness we again use the damping term to define a
small parameter $\epsilon=\nu$, and take $\Delta_{n } =
\epsilon\delta_{n}$, $a=\epsilon\alpha/3$, $D=\epsilon\beta$. The
oscillating displacement is then written as a slow modulation of
oscillations at frequency one, plus corrections
\begin{equation}
u_{n}(t)=\left[  A_{n}(T)e^{it} + c.c.\right]  + \epsilon
u_{n}^{(1)}(t) + \ldots\label{ansatz1}%
\end{equation}
with $T=\epsilon t$ a slow time scale\index{multiple scales!ansatz} as
before. As always, the slow variation of $A_{n}(T)$ gives us the extra
freedom to eliminate secular terms and ensure that the perturbative
correction $u_{n}^{(1)}(t)$, as well as all higher-order corrections
to the linear response, do not diverge. Note that our decision to
scale \eqref{eq:syncmodel} by setting the%
\index{van der Pohl oscillator} van der Pohl term such that the%
\index{nonlinear saturation} nonlinear saturation of the oscillations
occurs at $u_{n}=O(1)$, has affected the scaling of our trial solution
(\ref{ansatz1}), whose leading term is indeed of order 1. Compare this
with the trial solution of the previous section, given by
\eqref{eq:ansatz}, whose leading term is of $O(\sqrt\epsilon)$.

Using the relation (\ref{eq:adot}), again denoting a time derivative
with respect to the slow time $T$ by a prime, we calculate the time
derivatives of the trial solution~(\ref{ansatz1})
\begin{subequations}
\label{eq:derivs}%
\begin{align}
  \dot{u}_{n}  &  =\left(  [iA_{n}+\epsilon A_{n}^{\prime}]e^{it}+c.c.\right)
  +\epsilon\dot{u}_{n}^{(1)}(t)+\ldots\\
  \ddot{u}_{n}  &  =\left(  [-A_{n}+2i\epsilon A_{n}^{\prime}+\epsilon^{2}%
    A_{n}^{\prime\prime}]e^{it}+c.c.\right)  +\epsilon\ddot{u}_{n}^{(1)}(t)+\ldots
\end{align}
\end{subequations}
Substituting these expressions back into the scaled equation of motion
\index{equation of motion!one-dimensional array!scaled}
\begin{equation}\label{eq:scaledsyncmodel}
  \ddot{u}_{n} + \left(1+\epsilon\delta_n\right)u_{n} - \epsilon\left[\left(1-u_{n}^{2}\right)\dot{u}_{n} +
  \tfrac{1}{3} \alpha u_{n}^{3} + \tfrac{1}{2} \beta\left(u_{n+1}-2u_{n}+u_{n-1}\right)\right]
  = 0,
\end{equation}
and picking out all terms of order $\epsilon$, we get the following equation
for the first perturbative correction
\begin{align}
\ddot{u}_{n}^{(1)} + u_{n}^{(1)} = & -\delta_{n}A_{n}
-\left(2iA_{n}^{\prime}e^{it} + c.c. \right)  +\left(  iA_{n}e^{it} +
  c.c.\right)  \left[1 - \left(  A_{n}e^{it}+c.c.\right)^{2}\right]\nonumber\\ 
&  -\tfrac{1}{3} \alpha\left(  A_{n}e^{it}+c.c.\right)^{3} + \tfrac{1}{2} \beta \left[\left(  A_{n+1}
    - 2A_{n} + A_{n-1} \right)  e^{it} + c.c. \right]. \label{eq_xone}%
\end{align}

Terms varying as $e^{\pm3it}$ on the right-hand side of
Eq.~(\ref{eq_xone}) contribute a finite response to $u_{n}^{(1)}$, but
the collection of terms proportional to $e^{it}$---the secular
terms---act like a\index{secular term} force driving the simple
harmonic oscillator on the left-hand side at its resonance
frequency. The sum of all these secular terms must vanish so that the
perturbative correction $u_{n}^{(1)}(t)$ in \eqref{ansatz1} will not
diverge. This provides us with a \index{solvability condition}
solvability condition that leads to an equation for determining the
slowly varying amplitudes $A_{n}(T)$
\begin{equation}
  2\frac{dA_{n}}{dT} = (1 + i\delta_{n})A_{n}-(1 - i\alpha)
    \left\vert A_{n}\right\vert ^{2}A_{n} + i \frac{\beta}{2} \left(A_{n+1}
    - 2A_{n} + A_{n-1}\right) . \label{DuffingAmpA}%
\end{equation}
With a rescaling of time $\tau = T/2 = \epsilon t/2$ and a slight
rearrangement of terms, \eqref{DuffingAmpA} reduces to the form
obtained by Cross \emph{et al.}~\citeyear{sync1,sync2},
\index{amplitude equation!discrete}
\begin{equation}
  \boxed{
    \frac{dA_{n}}{d\tau} = i(\delta_{n} + \alpha \left\vert A_{n}\right\vert ^{2})A_{n} + (1 - 
    \left\vert A_{n}\right\vert ^{2})A_{n} + i \frac{\beta}{2} \left(A_{n+1}
    - 2A_{n} + A_{n-1}\right) 
  }\ . \label{SyncAmp}%
\end{equation}

The first term on the right-hand side shows the ability of the
$n^{th}$ oscillator to shift its frequency\index{frequency pulling} by
an amount $\alpha \left\vert A_{n}\right\vert^2$; the second term
shows the tendency of the oscillators to increase their amplitude as
long as $\left\vert A_{n}\right\vert^2 < 1$; and the third term is the
reactive coupling\index{coupling!linear reactive} between nearest
neighbors.  If this nearest-neighbor coupling is generalized to allow
also dispersive interaction and replaced by an all-to-all or
mean-field coupling, convenient for theoretical analysis, we obtain
the final form of the model studied by Cross \emph{et
  al.}~\citeyear{sync1,sync2}, 
\index{amplitude equation!discrete!mean-field}
\begin{equation}
  \boxed{
    \frac{dA_{n}}{d\tau} = i(\delta_{n} + \alpha \left\vert A_{n}\right\vert ^{2})A_{n} + (1 - 
    \left\vert A_{n}\right\vert ^{2})A_{n} + \frac{K+i\beta}{N}
    \sum_{m=1}^{N} \left(A_{m} - A_{n} \right) 
  }\ , \label{SyncAmpMF}%
\end{equation}
except for the fact that we use the opposite sign convention for the
Duffing parameter. Thus, in our current discussion a positive
(negative) value of $\alpha$ implies a stiffening (softening) Duffing
nonlinearity.%
\index{Duffing nonlinearity!stiffening}%
\index{Duffing nonlinearity!softening} The relative natural frequency
$\delta_{n}$ of each oscillator is chosen from a specified
distribution $g(\delta)$, whose width is denoted by $w$.

When only nonlinear saturation and\index{coupling!linear dissipative}
dissipative coupling are present ($\alpha=\beta=0,K\neq0$)
\eqref{SyncAmpMF} reduces to
\begin{equation}
\frac{dA_{n}}{d\tau} = (i\delta_{n} + 1 - \left\vert A_{n}\right\vert
^{2})A_{n} + \frac{K}{N}\sum_{m=1}^{N}(A_{m}-A_{n}),\label{AmplitudePhase}%
\end{equation}
which has been analyzed by \shortciteN{MMS91} for general $w$ and $K$.

\subsection{Analyzing and solving the equations}

The complex number $A_{n}$, representing the amplitude $r_{n}$ and
phase $\theta_{n}$ of the $n^{th}$ oscillator,
$A_{n}=r_{n}e^{i\theta_{n}}$, suggests the introduction of a
complex order parameter $\Psi$ to measure the coherence of the
oscillations\index{synchronization!order parameter}
\begin{equation}
  \Psi=R\,e^{i\Theta}=\frac{1}{N}\sum_{n=1}^{N}r_{n}e^{i\theta_{n}}.
  \label{OrderParameter}%
\end{equation}
A nonzero value of the order parameter $R>0$ may be taken as the
definition of a synchronized state.

The general amplitude-phase model reduces to familiar phase only
models of synchronization in certain limits.  If the width $w$ of the
distribution $g(\delta)$ is narrow, so that the time evolution of the
magnitudes $r_n=\left\vert A_{n}\right\vert $ is fast compared with
that of the phase dispersion, and the coupling constants $K,\beta$ are
small, $r_n$ rapidly relaxes to a value close to unity
\begin{equation}
r_{n}^{2}\simeq 1+\frac{K}{N}\sum_{m=1}^{N}[\cos(\theta_{m} -
  \theta_{n})-1] - \frac{\beta}{N}\sum_{m=1}^{N}\sin(\theta_{m} -
  \theta_{n}),
\end{equation}
and the only remaining dynamical variable for each
oscillator is its phase $\theta_{n}$. Equation~(\ref{SyncAmpMF})
can then be reduced to
\begin{equation}
  \dot{\theta}_{n}=\delta_{n}+\alpha+\frac{K-\alpha\beta}{N}\sum_{m=1}^{N}\sin(\theta_{m} -
  \theta_{n})+\frac{\alpha K+\beta}{N}\sum_{m=1}^{N}[\cos(\theta_{m} -
  \theta_{n})-1].\label{ExtendedKuramoto}
\end{equation}
For the case of purely dissipative coupling $\alpha=\beta=0,K\ne 0$,
or reactive coupling with strong frequency pulling
$K=0,\alpha,\beta\ne 0,|\alpha|\gg 1$, the last term on the right hand
side of eqn (\ref{ExtendedKuramoto}) can be neglected, and the
equation reduces to a simple form \shortcite{W67,K75}, known as the
Kuramoto model\index{Kuramoto model} (ignoring the unimportant
constant term $\alpha$ and writing the effective coupling constant in
either case simply as a $K$)
\begin{equation}\label{Kuramoto}
  \dot{\theta}_{n}=\delta_{n}+\frac{K}{N}\sum_{m=1}^{N}\sin(\theta_{m} -
  \theta_{n}),
\end{equation}
that has been the subject of numerous studies \shortcite{ABVRS05}. In
the absence of coupling each oscillator in this model would simply
advance at a rate that is constant in time, but with some dispersion
of frequencies over the different elements.

Identifying the imaginary part of $\Psi e^{-i\theta_{n}}$ in the sum
appearing in \eqref{Kuramoto}---while recalling that $r_{n}=1$ for the
Kuramoto model---yields a particularly simple mean-field expression
\begin{equation}
\dot{\theta}_{n}=\delta_{n}+KR\sin(\Theta-\theta_{n}). \label{KuramotoField}%
\end{equation}
Thus the behavior of each oscillator is given by its tendency to lock
to the phase of the\index{synchronization!order parameter} order
parameter. The term $KR\sin(\Theta-\theta_{n})$ acts as a locking
force, and locking occurs for all oscillators with frequencies
satisfying $\left\vert \delta_{n}\right\vert <KR$, with the locked
oscillator phase given by $\Theta+\sin^{-1}(\delta_{n}/KR)$. The
magnitude $R$ of the order parameter must then be determined
self-consistently via \eqref{OrderParameter}.

Equation~(\ref{Kuramoto}) is known to show rich behavior, including,
in the large $N$ limit, a sharp synchronization transition at some
value of the coupling constant $K=K_{c}$ \shortcite{K75}, which
depends on the frequency distribution $g(\delta)$ of the uncoupled
oscillators. The transition is from an unsynchronized state with
$\Psi=0$ in which the oscillators run at their individual frequencies,
to a synchronized state with $\Psi\neq0$ in which a finite fraction of
the oscillators lock to a single frequency. The transition at $K_{c}$
has many of the features of a second order phase transition, with
universal power laws and critical slowing down \shortcite{K75}, as
well as a diverging response to an applied force \shortcite{S88}.

The last term in eqn (\ref{ExtendedKuramoto}) may lead to important
qualitative effects even if the coefficient is not very large. For
example, in the absence of this term the coupling terms cancel when
summing over all the oscillators in the system, so that the frequency
of a synchronized state is simply related to the mean frequency of the
oscillators. This is no longer the case for the general equation.  In
the case of short range, rather than all to all coupling, the
$\cos(\theta_{m}-\theta_{n})$ term profoundly changes the nature of
the synchronized state to one of propagating
waves~\shortcite{sakaguchi88b,blasius05}.

\subsection{Brief survey of applications}

\begin{figure}[tb]
\centering
\subfigure{\includegraphics[width=0.46\textwidth,height=0.385\textwidth]{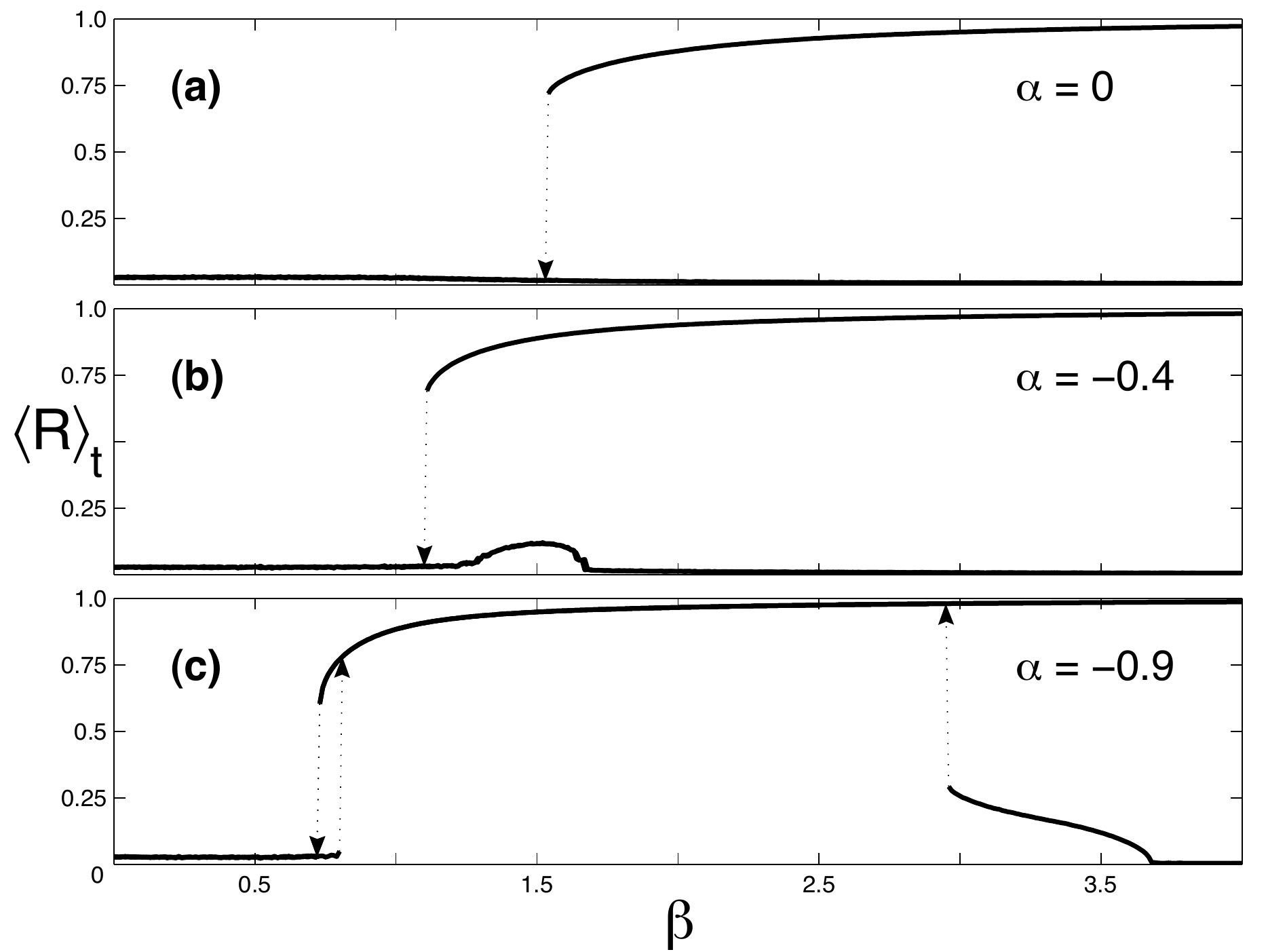}}
\subfigure{\includegraphics[width=0.46\textwidth,height=0.4\textwidth]{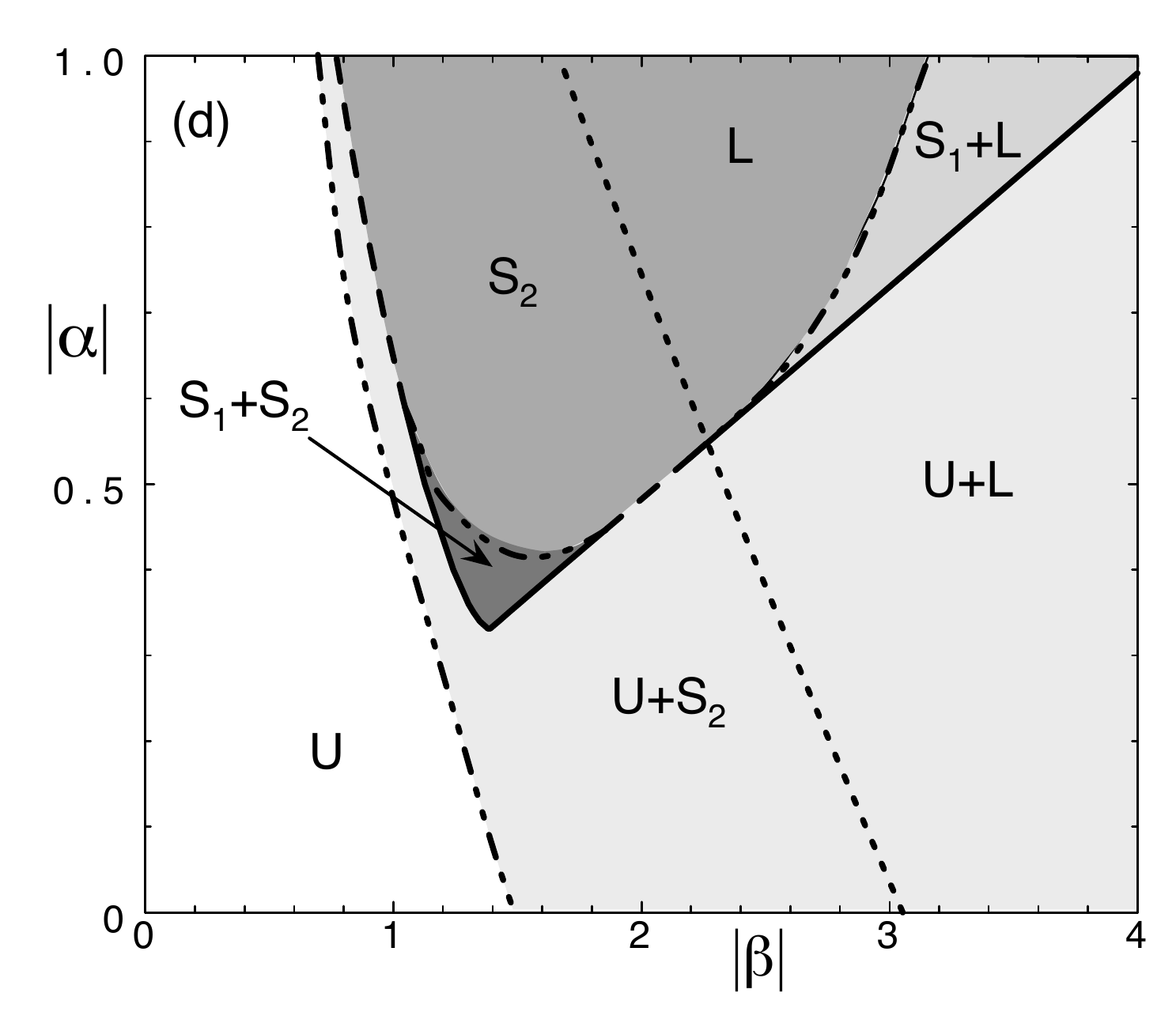}}
\caption{\label{fig:sync}(a)-(c) Simulations of $1000$ oscillators
  having a triangular frequency distribution with width $w=2$. The
  time-averaged order parameter magnitude $\langle R\rangle_{t}$ is
  plotted for both upward and downward sweeps of $\beta$ at fixed
  $\alpha$: (a) $\alpha=0.0$; (b) $\alpha=-0.4$; and (c)
  $\alpha=-0.9$. The same results would be obtained by switching the
  signs of both $\alpha$ and $\beta$. (d) Stability
  diagram\index{stability diagram} for the same triangular
  distribution, in the quadrant of the $\alpha-\beta$ plane with
  $\alpha\beta<0$. Solid and dashed lines show analytical results of
  the linear stability of the unsynchronized state. Numerics show the
  bifurcations are supercritical\index{bifurcation!supercritical}
  along the solid portions and
  subcritical\index{bifurcation!subcritical} along the dashed
  portion. Dotted line is the linear stability boundary of the fully
  locked state.  Dash-dotted lines are saddle-node bifurcations
  observed in numerical simulations. States are: $U$ - unsynchronized;
  $S_{1},S_{2}$ synchronized with small and large amplitude
  respectively; $L$ fully locked. From
  Cross \emph{et al.} (2006). Copyright (2006) American Physical Society.}
\end{figure}

The synchronization of oscillators with reactive all-to-all coupling
and nonlinear frequency pulling, described by of \eqref{SyncAmpMF}
with $K=0$, was analyzed by Cross \emph{et al.}~\citeyear{sync1,sync2}
for several different frequency distributions $g(\delta)$ (Lorentzian,
top-hat, and triangular). Here we briefly review the results for a
triangular distribution with width $w=2$, and refer the reader to the
original work for more details. Such a width is not small compared
with the relaxation rate of the magnitude variables, and so the
behavior is richer than in the weak randomness limit described by the
Kuramoto model. The stability diagram\index{stability diagram} of the
variety of states found as $\alpha$ and $\beta$ are varied is shown in
Fig.~\ref{fig:sync}(d). The results are shown for $\alpha\beta<0$,
noting that for a symmetric distribution of frequencies the results
are the same if both signs of $\alpha$ and $\beta$ are changed. These
same results were presented by Cross \emph{et
  al.}~\citeyear{sync1,sync2} for their case $\alpha\beta>0$, as they
were using the opposite sign convention for $\alpha$.

 Certain results can be obtained analytically, in particular the
instability from the unsynchronized state to a synchronized state with
nonzero order parameter $R>0$, and the instability from the fully
locked state (all oscillators locked to evolve at the same frequency)
to a synchronized state with only partial locking. Other results are
obtained numerically by performing sweeps of $\beta$ at fixed values
of $\alpha$. Results for the time averaged magnitude of the order
parameter $\langle R\rangle_t$ for three values of $\alpha$ are shown in
Fig.~\ref{fig:sync}(a)-(c). Both upward and downward quasistatic
sweeps\index{quasistatic sweep!of coupling strength} of the reactive
coupling strength $\beta$ are used to uncover
hysteresis\index{hysteresis} in the transitions.

Many novel features are apparent in these results. For example, the
unsynchronized state is stable for both small and large values of the
coupling strength $\beta$, so that for fixed $\alpha$ there are
\emph{two} values of $\beta$ at which the unsynchronized state passes
from stable to unstable. At large values of $\beta$ a large
order-parameter synchronized state is also stable, which becomes the
fully locked state at large enough $\beta$. The transition from the
unsynchronized state to the synchronized state may be continuous,
passing through a supercritical bifurcation, or discontinuous, passing
through a subcritical bifurcation. Hysteresis is apparent in the
latter case, owing to the bistability\index{multistability} of both
the synchronized and the unsynchronized states. This is demonstrated
explicitly in Fig.~\ref{fig:sync}(c). More surprisingly, one also
observes the multistability of different synchronized states. Over
some parameter ranges a small order-parameter synchronized state may
coexist with the large order-parameter synchronized state, as
observed, for example, in Fig.~\ref{fig:sync}(c) between about
$\beta=3.0$ and $\beta=3.7$. This small order-parameter synchronized
state has the novel property that the order parameter is nonzero
$R>0$, but there is \emph{no} oscillator locked in frequency to the
frequency of the order parameter or of other oscillators---this is a
\emph{synchronized} state $R>0$ with \emph{no frequency locking}.

The rich synchronization behavior displayed by this stability diagram
opens up many possibilities for applications, as well as suggesting
difficulties that must be overcome, for example when there exists a
multistability of different dynamical states.

\index{synchronization|)}

\section{Continuous Amplitude Equations:\\
Example III -- Nonlinear competition between extended modes}
\label{sec:patterns}

\subsection{Derivation of the BCL amplitude equation}
\index{BCL equation|(}

We wish to investigate the sequence of single mode standing
wave patterns to be expected in parametrically driven resonator
arrays, in cases where many such modes are\index{multistability}
simultaneously stable, when the strength of the drive is
varied. Although the quantitative analysis could be done directly from
the basic equations of motion for the coupled resonators, it is
advantageous to formulate the analysis in terms of a continuous
amplitude equation---that which was developed by \citeN{BCL},
henceforth referred to as the BCL equation. This allows us to display
the range of stable patterns on a reduced \index{stability diagram}
\emph{stability diagram} involving just two dimensionless variables (a
scaled measure of the driving strength, and a scaled mode wave
number), so that it is easy to deduce the general qualitative behavior
upon variation of the parameters. The specific quantitative behavior
for a physical system is also easy to obtain by evaluating the
corresponding scaled quantities. A change of pattern occurs when
parameters vary so that the mode moves outside of the region of stable
patterns on this diagram, and the new pattern is predicted by
analyzing the result of the instability using the BCL equation. This
type of approach was used in other pattern forming systems
\shortcite{ksz}. A novel feature of the present system is that the
difference in the instabilities encountered on increasing and
decreasing the (scaled) driving strength leads to the prediction of
quite different-sized mode jumps for the up and down sweeps.

We follow the treatment of BCL in deriving their amplitude equation,
but instead of starting with the original equations of
motion~(\ref{LCeom}) derived by \citeN{LC}, we start with the simpler
equations of motion~(\ref{eq:eom}). This leads to a somewhat
simplified derivation, which eventually yields the same amplitude
equation to describe the slow dynamics of the system of resonators. We
perform the same scaling of the equation parameters as we did in
\secref{sec:LC}, with one difference---in anticipation of treating
extremely large arrays, with thousands or more normal modes of
vibration, we do not wish to assume that the width of the frequency
band is small. We therefore \emph{do not} replace $D$ with $\epsilon
d$ as before. This will also allow us to obtain the exact dispersion
relation~(\ref{eq:arraydispersion}) at the linear step, \emph{i.e.} at
order $\sqrt\epsilon$.  Our starting point is therefore the set of
coupled equations
\index{equation of motion!one-dimensional array!scaled}
\begin{equation}
  \label{eq:bcleom}
  \ddot{u}_{n} + \epsilon\dot{u}_{n} + \left(1 - \epsilon
  h\cos2\omega_{p}t\right)u_{n} 
  + \tfrac{1}{2} D(u_{n+1}-2u_{n}+u_{n-1}) + u_{n}^{3} 
  + \eta u_{n}^{2}\dot{u}_{n}=0,
\end{equation}
where we have taken a negative sign for the parametric driving term to
be consistent with the sign used by BCL, thus merely shifting the
phase of the drive by $\pi$ relative to \eqref{eq:eom}.

\subsubsection{Amplitude Equations for Counter Propagating Waves}

In order to treat this system of equations analytically, beyond the
treatment described earlier in \secref{sec:LC}, we introduce a
continuous displacement field $u(x,t)$, keeping in mind that only for
integral values $x=n$ of the spatial coordinate does it actually
correspond to the displacements $u(n,t)=u_n(t)$ of the discrete set of
resonators in the array. We introduce slow spatial and temporal
scales, $X=\epsilon x$ and $T=\epsilon t$, upon which the dynamics of
the envelope function occurs, and expand the displacement field in
terms of $\epsilon$,\index{multiple scales!ansatz}
\begin{eqnarray} \label{uAnsatz}\nonumber
u(x,t)&=& \epsilon^{1/2}
\left[\left(A_+(X,T)e^{-iq_px}+A_-^*(X,T)e^{iq_px}\right)e^{i\omega_p
t} + c.c.\right]\\ &+& \epsilon^{3/2}u^{(1)}(x,t,X,T)+\ldots,
\end{eqnarray}
where the asterisk and $c.c.$ stand for the complex conjugate, and
$q_p$ and $\omega_p$ are related through the dispersion
relation,\index{dispersion relation}
\begin{equation}
  \label{eq:dispersion-qp}
  \omega_p^2=1-2D\sin^2\frac{q_p}{2}.
\end{equation}
Note that the response to lowest order in $\epsilon$ is expressed in
terms of two counter-propagating waves with complex amplitudes $A_+$
and $A_-$, which is a typical ansatz for parametrically excited
continuous systems~\cite{reviewcross}. We substitute the
ansatz~(\ref{uAnsatz}) into the equations of motion~(\ref{eq:bcleom})
term by term. Again, using \eqref{eq:adot} in addition to expanding
$A_\pm(X+\epsilon,T) \simeq A_\pm(X,T) + \epsilon \partial
A_\pm(X,T)/\partial X$ we obtain up to order $\epsilon^{3/2}$,

\begin{subequations}
\begin{align}
\ddot u_n&= \epsilon^{1/2} \left[\left(-\omega_p^2 A_+
    +2i\omega_p\epsilon  \frac{\partial A_+}{\partial T}\right) e^{-iq_px}
+ \left(-\omega_p^2 A_-^* +2i\omega_p\epsilon \frac{\partial A_-^*}{\partial
T}\right) e^{iq_px} \right] e^{i\omega_pt} \nonumber\\
& + c.c. +\epsilon^{3/2}\frac{\partial^2 u^{(1)}}{\partial t^2},\\
u_{n\pm1}&=\epsilon^{1/2} \left[
\left(A_+ \pm\epsilon \frac{\partial
A_+}{\partial X}\right)e^{-iq_p(x\pm1)} + \left(A_-^* \pm\epsilon \frac{\partial
A_-^*}{\partial X}\right) e^{iq_p(x\pm1)}\right] e^{i\omega_p
t} +c.c. \nonumber\\
&+\epsilon^{3/2}u^{(1)}(x\pm1,t,X,T),\\
\displaybreak[0]
\frac12 D&\left(u_{n+1}-2u_n+u_{n-1}\right)=-\epsilon^{1/2}2 D
\sin^2(q_p/2)\left(A_+e^{-iq_px}+A_-^*e^{iq_px}\right)e^{i\omega_p
t}\nonumber\\&-\epsilon^{3/2} i D \sin(q_p)\left(\frac{\partial
A_+}{\partial X}e^{-iq_px}-\frac{\partial A_-^*}{\partial
X}e^{iq_px}\right)e^{i\omega_p t}+c.c.\nonumber\\
&+\epsilon^{3/2} \frac{D}2 \left[u^{(1)}(x+1,t,X,T)-2u^{(1)}(x,t,X,T)+u^{(1)}(x-1,t,X,T)
\right],\\
\displaybreak[0]
\epsilon h{\rm c}&{\rm os}(2\omega_pt) u_n = \epsilon^{3/2} \frac{h}2
\left(A_-e^{-iq_px}+A_+^*e^{iq_px} 
\right)e^{i\omega_pt}+O(e^{i3\omega_pt})+c.c.,\\
\displaybreak[0]
\epsilon\dot u_n &= \epsilon^{3/2} i\omega_p
\left(A_+e^{-iq_px}+A_-^*e^{iq_px}\right) e^{i\omega_p t}+c.c.,
\\
\displaybreak[0]
u_n^3 &= \epsilon^{3/2} 3\left[\left(|A_+|^2+2|A_-|^2\right) A_+ e^{-iq_px}
+ \left(2|A_+|^2+|A_-|^2\right) A_-^*e^{iq_px}\right] e^{i\omega_p t}\nonumber\\
&+ O\left(e^{i3\omega_p t},e^{i3q_px}\right)+c.c.,
\displaybreak[0]
\end{align}
and
\begin{align}
 u_n^2\dot u_n
&= \epsilon^{3/2} i\omega_p \left[\left(|A_+|^2+2|A_-|^2\right) A_+ e^{-iq_px}
+ \left(2|A_+|^2+|A_-|^2\right) A_-^*e^{iq_px}\right] e^{i\omega_p t}\nonumber\\
&+ O\left(e^{i3\omega_p t},e^{i3q_px}\right)+c.c.,
\displaybreak[0]
\end{align}
\end{subequations}
where $O(e^{i3\omega_p t},e^{i3q_px})$ are fast oscillating terms
proportional to $e^{i3\omega_p t}$ or $e^{i3q_px}$ that do not enter
the dynamics at the lowest order in $\epsilon$ because they are
nonsecular.

At the order of $\epsilon^{1/2}$, the equations of
motion~(\ref{eq:bcleom}) are satisfied trivially, yielding the
dispersion relation~(\ref{eq:dispersion-qp}) mentioned earlier. At the
order of $\epsilon^{3/2}$ on the other hand, we again obtain secular
terms,\index{secular term} and must apply a \emph{solvability
  condition}, which requires that all terms proportional to
$e^{i(\omega_pt\pm q_px)}$ must vanish. As a result, we obtain the two
coupled amplitude equations,%
\index{amplitude equation!continuous} \index{solvability condition}
\begin{equation}\label{AmpEqs}
\frac{\partial A_\pm}{\partial T} \pm v_g\frac{\partial A_{\pm}}{\partial X} = -
\frac12 A_{\pm} \mp \frac{i h}{4\omega_p} A_{\mp}
- \frac12 \left(\eta \mp \frac{3i}{\omega_p}\right)
\left(|A_{\pm}|^{2}+2|A_{\mp}|^{2}\right)A_{\pm},
\end{equation}
where the upper signs (lower signs) give the equation for $A_+$
($A_-$), from the restriction on the terms proportional to
$e^{i\omega_pt-iq_px}$ ($e^{i\omega_pt+iq_px}$), and where
\begin{equation}
  \label{eq:vgroup}
  v_g =
\frac{\partial\omega}{\partial q} =
-\frac{D\sin(q_p)}{2\omega_p}
\end{equation}
is the group velocity.\index{group velocity} A detailed derivation of
the amplitude equations~(\ref{AmpEqs}) can be found in the Masters
thesis of~\shortciteN{yaron}.  Similar equations were previously
derived for describing Faraday waves~\shortcite{ezerskii,milner}.

\subsubsection{Reduction to a Single Amplitude Equation}

By linearizing eqns~(\ref{AmpEqs}) about the zero solution
($A_+=A_-=0$) we find that the linear combination of the two
amplitudes that first becomes unstable at $h_c=2\omega_p$ is $\hat
B\propto (A_+-iA_-)$---representing the emergence of a standing wave
with a temporal phase of $\pi/4$ relative to the drive---while the
orthogonal linear combination of the amplitudes decays exponentially
and does not participate in the dynamics at onset.  Thus, just above
threshold we can reduce the description of the dynamics to a single
amplitude $\hat B$, where at a finite distance above threshold
a band of unstable modes around $q_p$ can contribute to the spatial
form of $\hat B$. This is similar to the procedure introduced by
\citeN{riecke} for describing the onset of Faraday waves.

To proceed with our multiple scales analysis, and obtain an equation
describing the relevant slow dynamics of the new amplitude $\hat B$,
we need to identify another physically small parameter with which we
can associate even slower spatial and temporal scales. We therefore
assume that the coefficient of nonlinear
damping\index{damping!nonlinear} $\eta$ is small, and define a second
small parameter $\delta=\eta^2\ll 1$. We then define a reduced driving
amplitude $\hat g$ with respect to the threshold $h_c$ by letting
$(h-h_c)/h_c \equiv \hat g\delta$.  A sequence of judicious
arguments~\shortcite{yaron,BCL} then encourages us to scale the
original amplitudes $A_\pm$ as $\delta^{1/4}$, making the ansatz that%
\index{multiple scales!ansatz}
\begin{equation}\label{Bansatz}
  \left(%
    \begin{array}{c}
      A_+ \\
      A_- \\
    \end{array}%
  \right) = \delta^{1/4}
  \left(%
    \begin{array}{c}
      1 \\
      i \\
    \end{array}%
  \right) \hat{B}(\hat\xi,\hat\tau)+
  \delta^{3/4}\left(%
    \begin{array}{c}
      w^{(1)}(X,T,\hat\xi,\hat\tau) \\
      v^{(1)}(X,T,\hat\xi,\hat\tau) \\
    \end{array}%
  \right) +
  \delta^{5/4}\left(%
    \begin{array}{c}
      w^{(2)}(X,T,\hat\xi,\hat\tau) \\
      v^{(2)}(X,T,\hat\xi,\hat\tau) \\
    \end{array}%
  \right)+\ldots,
\end{equation}
where $\hat\xi=\delta^{1/2} X$ and $\hat\tau=\delta T$ are the new
spatial and temporal scales respectively. 

We substitute the
ansatz~(\ref{Bansatz}) into the coupled amplitude
equations~(\ref{AmpEqs}) and collect terms of different orders in
$\delta$. Again, to the lowest order of expansion the equations are
satisfied trivially. Collecting all terms of order $\delta^{3/4}$ in
eqns~(\ref{AmpEqs}) yields
\begin{equation}\label{B3_4order}
\mathfrak{O}\left(%
\begin{array}{c}
  w^{(1)} \\
  v^{(1)} \\
\end{array}%
\right)=\left(-v_g\frac{\partial\hat{B}}{\partial
\hat\xi}+i\frac9{2\omega_p}|\hat{B}|^2\hat{B}\right)
\left(%
\begin{array}{c}
  1 \\
  -i \\
\end{array}%
\right),
\end{equation}
where exactly at onset $\mathfrak{O}$ is a linear operator given by the matrix
\begin{equation} \label{B_Operator}
\left(%
\begin{array}{cc}
  \partial_T+v_g\partial_X+\frac12 & \frac{i}2 \\
  -\frac{i}2 & \partial_T-v_g\partial_X+\frac12 \\
\end{array}%
\right).
\end{equation}
The vector $\scriptsize\left(%
\begin{array}{c}
  1 \\
  -i \\
\end{array}%
\right)$, on the right hand side of \eqref{B3_4order} is an
eigenvector of $\mathfrak{O}$, with an eigenvalue $-1$. The solution
of \eqref{B3_4order} is therefore immediately given by
\begin{equation}\label{w1v1Sol}
\left(\begin{array}{c}
  w^{(1)} \\
  v^{(1)} \\
\end{array}%
\right) = \left(-v_g\frac{\partial\hat{B}}{\partial \hat\xi} +
  i\frac9{2\omega_p}|\hat{B}|^2\hat{B}\right) \left( 
\begin{array}{c}
  1 \\
  -i \\
\end{array}%
\right).
\end{equation}
We substitute \eqref{w1v1Sol} back into eqns~(\ref{AmpEqs}), collect
all the terms of order $\delta^{5/4}$ and obtain
\begin{multline}\label{B5_4order}
\mathfrak{O}\left(%
\begin{array}{c}
  w^{(2)} \\
  v^{(2)} \\
\end{array}%
\right) = \left[-\frac{\partial\hat{B}}{\partial\hat\tau} + v_g^2
  \frac{\partial^2\hat{B}}{\partial\hat\xi^2} + \frac{\hat{g}}2 \hat{B} - \frac32
  |\hat{B}|^2\hat{B} \right.\\ \left.
-i\frac{3 v_g}{\omega_p} \left(4|\hat{B}|^2\frac{\partial
    \hat{B}}{\partial\hat\xi} 
+ \hat{B}^2\frac{\partial \hat{B}^*}{\partial\hat\xi}\right) 
- \left(\frac{9}{2\omega_p}\right)^2 |\hat{B}|^4\hat{B}\right]
\left(%
\begin{array}{c}
  1 \\
  i \\
\end{array}%
\right).
\end{multline}
The vector $\scriptsize\left(%
\begin{array}{c}
  1 \\
  i \\
\end{array}%
\right)$, on the right hand side of \eqref{B5_4order}, is an
eigenvector of $\mathfrak{O}$ with zero eigenvalue. Clearly, the left
hand side of the equation cannot contain any component along the
direction of such a zero eigenvector. Therefore, the expression within
the square brackets is a\index{secular term} secular term that must
vanish. This provides us with the required \emph{solvability
condition}\index{solvability condition} to proceed~\cite{reviewcross}.
After applying one last set of rescaling transformations,
\begin{equation} \label{scaling}
\hat\tau = \frac{36}{\omega_p^2} \tau, \quad
\hat\xi = \frac{6 |v_g|}{\omega_p}\xi, \quad
\hat{B} = \frac{\omega_p}{3\sqrt{3}} B, \quad {\rm and}\quad
\hat{g} = \frac{\omega_p^2}{18} g,
\end{equation}
we end up with the BCL amplitude equation,\index{BCL equation|textbf}
which is governed by a single parameter,%
\index{amplitude equation!continuous}
\begin{equation} \label{BampEq}
  \boxed{
  \frac{\partial B}{\partial\tau} = gB +
  \frac{\partial^{2}B}{\partial\xi^{2}} + i\frac{2}{3}
  \left(4|B|^{2}\frac{\partial B}{\partial\xi} +
    B^{2}\frac{\partial B^{*}}{\partial\xi}\right)
  - 2|B|^{2}B - |B|^{4}B
  }\ .
\end{equation}

\subsection{Analyzing and solving the equation}

The simplest nontrivial solutions of the BCL amplitude
equation~(\ref{BampEq}) are steady-state single-mode extended
patterns, given by\index{steady-state solution!scaled}%
\index{single-mode extended solution!scaled}
\begin{equation}\label{SingleMode}
B(\xi,\tau) = b_k e^{i(\varphi-k\xi)},
\end{equation}
with $b_k$ and $\varphi$ both real, and where the boundary conditions
$u(0,t)=u(N+1,t)=0$\index{boundary conditions!fixed} constrain the
phase $\varphi$ to be $\pi/4$ or $5\pi/4$.  In steady state, the
relation between the magnitude $b_k$ and the wave number $k$ is found
by substituting \eqref{SingleMode} into \eqref{BampEq}, and setting
the time derivative to zero to give
\begin{equation}\label{Bsteady}
b_k^2 = (k-1) + \sqrt{(k-1)^2 + (g-k^2)}\geq 0,
\end{equation}
along with a negative square-root branch which is always unstable
against small perturbations, as can be verified by the analysis below.

Substituting the single-mode solution of \eqref{SingleMode}, with
$\varphi=\pi/4$, back into \eqref{w1v1Sol} and \eqref{Bansatz}, and
then into \eqref{uAnsatz}, yields extended single-mode standing-wave
parametric oscillations at half the drive frequency, whose explicit
form is given by\index{steady-state solution}%
\index{single-mode extended solution}
\begin{eqnarray}\label{second_order}
  u(x,t) &\simeq &\epsilon^{1/2}\delta^{1/4} 
  \frac{4\omega_p \sqrt{1 + \tan^{2}(\alpha)}}{3\sqrt{3}} 
  b_{k} \sin(q_{m}x) \cos(\pi/4-\omega_{p}t-\alpha),
\end{eqnarray}
where we have defined
\begin{equation}
  \label{eq:tanalpha}
  \tan(\alpha) \equiv \delta^{1/2} \frac{\omega_{p}}{6}
  \left(b_{k}^2 - k\right).
\end{equation}
To satisfy the boundary conditions $u(0,t)=u(N+1,t)=0$, the wave
\index{boundary conditions!fixed} numbers $q_m$ must be of the form
\begin{equation}
\label{eq:q-quantization}
 q_{m}=\frac{m\pi}{N+1}=q_{p}+\frac{k\pi}{\Delta Q_{N}(N+1)},
\end{equation}
where
\begin{equation}
  \label{eq:Qmin}
  \Delta Q_N = \frac{1}{\epsilon\delta^{1/2}}
  \frac{3D\sin(q_p)}{\omega_{p}^2} \frac{\pi}{N+1}.
\end{equation}

BCL showed that the first single-mode pattern to emerge as the
zero-state becomes unstable is the one whose wave number $q_m$ is closest
to the wave number $q_p$ that is determined by the drive frequency
$\omega_p$ through the dispersion relation~(\ref{eq:dispersion-qp}). This
determines the value of the scaled wave number in the single-mode
solution~(\ref{SingleMode}) to be
\begin{equation}\label{eq:k-zero}
k_0 =  \left(m - q_p \frac{N+1}{\pi}\right) \Delta Q_N,
\end{equation}
where $m$ is the integer closest to $q_p(N+1)/\pi$. Note that $\Delta
Q_N$ tends to zero as the size $N$ of the array of resonators tends to
infinity.  

\begin{figure}[b]
\centering
\includegraphics[width=0.7\columnwidth]{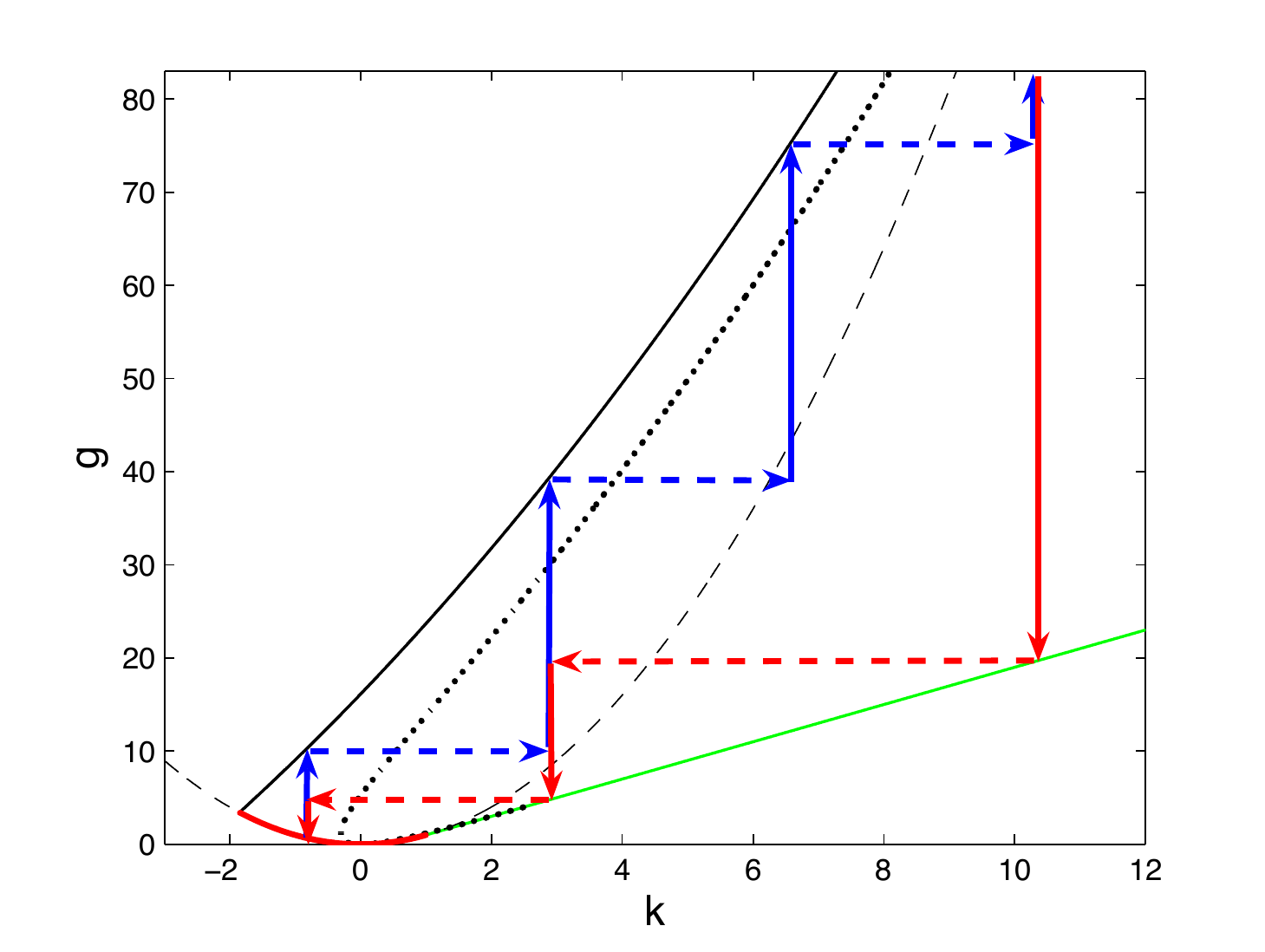}%
\caption{\label{StbBalloon} (Color) Stability boundaries of the single-mode
  solution (\ref{SingleMode}) of the BCL amplitude equation
  (\ref{BampEq}) in the $g$~{\it vs.}~$k$ plane.  Dashed line: neutral
  stability curve \index{neutral stability curve} $g=k^{2}$. Dotted
  line: stability boundary of the single-mode
  solution~(\ref{SingleMode}) for a continuous spectrum ($\Delta
  Q_N\rightarrow0$).  Solid lines: stability boundary of the
  single-mode solution for $N=92$ and the parameters $D=0.25$,
  $q_{p}=73\pi/101$, and $\epsilon=\delta = 0.01$ (giving $k_0\simeq
  -0.81$ and $\Delta Q_N \simeq 3.70$).  Black line: the value of $g$
  for which perturbations of the form given by \eqref{BStability}
  start to grow.  Red line: the lower bound for $k<1$, $g=k^{2}$.
  Green line: the lower bound for $k>1$, the locus of saddle-node
  bifurcations $g=2k-1$.  Vertical and horizontal arrows mark the
  secondary instability transitions that are expected upon quasistatic
  sweeps of $g$.  \index{quasistatic sweep!of drive amplitude} The
  blue upward-pointing arrows are for upward sweeps that undergo an
  Eckhaus instability, and the red downward-pointing arrows are for
  downward sweeps, of which the two with $k>1$ experience a
  saddle-node bifurcation, and the one with $k<1$ goes through a
  continuous (supercritical) bifurcation. From Kenig \emph{et al.}
  (2009$a$). Copyright (2009) American Physical Society. \index{stability diagram}}
\end{figure}

\index{single-mode extended solution!stability|(}

Linearization of the BCL amplitude equation (\ref{BampEq}) shows that
the zero state with $B(\xi,\tau)=0$---which is a solution
of~\eqref{BampEq} for any value of $g$---is stable against the
formation of single-mode patterns with wave number $k$ as long as
$g<k^2$. The neutral stability\index{neutral stability curve} curve
$g=k^2$ is plotted as a dashed parabola in
Fig.~\ref{StbBalloon}. Furthermore, for $k<1$ the bifurcation from the
zero state to that of single-mode oscillations is
supercritical,\index{bifurcation!supercritical} occurring on the
neutral stability curve, while for $k>1$ it is
subcritical,\index{bifurcation!subcritical} with a locus of
saddle-node bifurcations\index{bifurcation!saddle-node} located along
the line $g=2k-1$ (shown in Fig.~\ref{StbBalloon} as a solid green
line), where the square root in \eqref{Bsteady} is exactly zero.

The stability of a single-mode solution (\ref{SingleMode}) of wave
number $k$ against an Eckhaus transition\index{Eckhaus instability} to
a different single-mode solution of wave number $k\pm Q$ is found by
performing a linear stability analysis of solutions of the form
\begin{equation}\label{BStability}
B(\xi,\tau)=b_ke^{-ik\xi}+\left(\beta_+(\tau)
e^{-i(k+Q)\xi}+\beta_-^*(\tau)e^{-i(k-Q)\xi}\right),
\end{equation}
with $|\beta_\pm|\ll1$. When the larger of the two eigenvalues
describing the growth of such a perturbation becomes positive the
single-mode solution of wave number $k$ undergoes an Eckhaus
instability with respect to single-mode solutions of wave
numbers $k\pm Q$.

For an infinite number of oscillators the Eckhaus
instability\index{Eckhaus instability} forms the upper boundary of the
stability balloon of the single-mode solutions, and also the lower
boundary for $k<5/2$. For $k>5/2$ the lower boundary is the saddle
node bifurcation\index{bifurcation!saddle-node} line. For a finite
number of oscillators, restricting $Q$ to be an integer multiple of
$\Delta Q_N$ in \eqref{BStability} slightly shifts the Eckhaus
instability lines. The upper Eckhaus boundary is shifted to larger
values of $g$. The nature of the lower instability boundary now
depends on the number of resonators in the array through $\Delta
Q_{N}$, as well as on the wave number $k$. For $k<1$ the lower
boundary will be the Eckhaus instability curve if $|k|>\Delta
Q_{N}/2$, and the neutral stability curve otherwise.  Because the only
wave number to satisfy $|k|<\Delta Q_{N}/2$ is $k_{0}$, given by
\eqref{eq:k-zero}, upon decreasing $g$ the $k_{0}$ solution undergoes
a continuous transition to the zero state. For $k>1$ the lower
boundary will be the Eckhaus instability curve if $1<k<(5-3(\Delta
Q_{N}/2)^{2})/2$, and the line of saddle node bifurcations
otherwise. For $\Delta Q_{N}>2$ there is no portion of Eckhaus
instability\index{Eckhaus instability} on the lower boundary, which is
the neutral stability curve if $k<1$ and the saddle node bifurcation
curve if $k>1$. These stability boundaries are shown in
Fig.~\ref{StbBalloon} for an infinite system and for a system of
$N=92$ resonators. Further details can be found in \shortciteN{BCL}
and \shortciteN{kenig09}.  

\index{single-mode extended solution!stability|)}

\subsection{Brief survey of applications}

\shortciteN{kenig09} used the BCL amplitude equation~(\ref{BampEq}) to
study a number of collective dynamical effects in one-dimensional
arrays of coupled nonlinear resonators. The common thread linking
these effects is the so-called question of \emph{pattern
  selection}\index{pattern selection} \cite{crossBook}---the nonlinear
competition between different single-mode standing-wave patterns of
the form of \eqref{second_order}, when many such solutions are
simultaneously stable.\index{multistability} This question becomes
particularly interesting when the control parameter---in our case the
drive amplitude $g$---is varied as a function of time, either
quasistatically, abruptly, or in an intermediate ramp rate. In all
such cases one is interested in predicting which of all stable
patterns will be selected, as well as in the detailed understanding of
the nature of the switching between patterns as their stability
changes.

The BCL amplitude equation allowed \shortciteN{kenig09} to map out the
expected behavior of the resonators using universal stability
diagrams,\index{stability diagram} like the one shown in
Fig.~\ref{StbBalloon}. Such a diagram immediately shows the type of
instability that will be encountered upon variation of the control
parameter, and gives qualitative insights on the mode jumps to be
expected. For example, for\index{quasistatic sweep!of drive amplitude}
quasistatic parameter variations the jump in the mode number is always
unity if the control parameter is increased so that the Eckhaus
instability operates, but larger jumps are often seen if the control
parameter is decreased so that a saddle-node bifurcation occurs. This
is indicated by the dashed sideways arrows in Fig.~\ref{StbBalloon}.

It is instructive to describe how \shortciteN{kenig09} examined the
process by which these two types of pattern switchings occur. To do
so, we expand the general solution of the BCL amplitude equation in
the linear modes of the array
\begin{equation}\label{multimode}
  B(\xi,\tau)=\sum_n b_{n}(\tau)e^{i(\varphi_{n}-k_{n}\xi)},
\end{equation}
where $k_{n}\equiv k_0 + n\Delta Q_N$, and $k_0$ is defined in
\eqref{eq:k-zero}. Substituting a truncated mode
expansion~(\ref{multimode}) containing a finite number of modes around
$k_0$ into the BCL amplitude equation~(\ref{BampEq}), allows us to
replace this partial differential equation with a finite number of
ordinary differential equations for the coupled mode amplitudes,
\begin{eqnarray}\label{a-eqn}\nonumber
\frac{\partial b_{n}}{\partial \tau}&= &\left(g - k_{n}^{2}\right)
 b_{n}
 + 2\sum_{m,p}\left(k_p - 1 - \frac{m-n}{3} \Delta Q_N \right)
 b_{m}b_{p}b_{m+p-n}^{*}\\
 &- &\sum_{m,l,p,r}b_{m}b_{l}^{*}b_{p}b_{r}b_{m-l+p+r-n}^{*}.
\end{eqnarray}
To satisfy the boundary conditions, as mentioned above for the
single-mode solution~(\ref{SingleMode}), we take each mode amplitude
to be zero at the boundaries by setting all the phases $\varphi_{n}$
in Eq.~(\ref{multimode}) to $\pi/4$, and take the amplitudes $b_{n}$
to be real, keeping in mind that they can be either positive or
negative.  Note that if all mode amplitudes except $b_0$ are set to
zero we obtain a single equation with $n=m=p=l=r=0$, whose
steady-state solution is the same as the single-mode solution of
BCL~(\ref{Bsteady}).

\begin{figure}[b]
\centering
\subfigure[Eckhaus instability]{\includegraphics[width=0.5\textwidth]{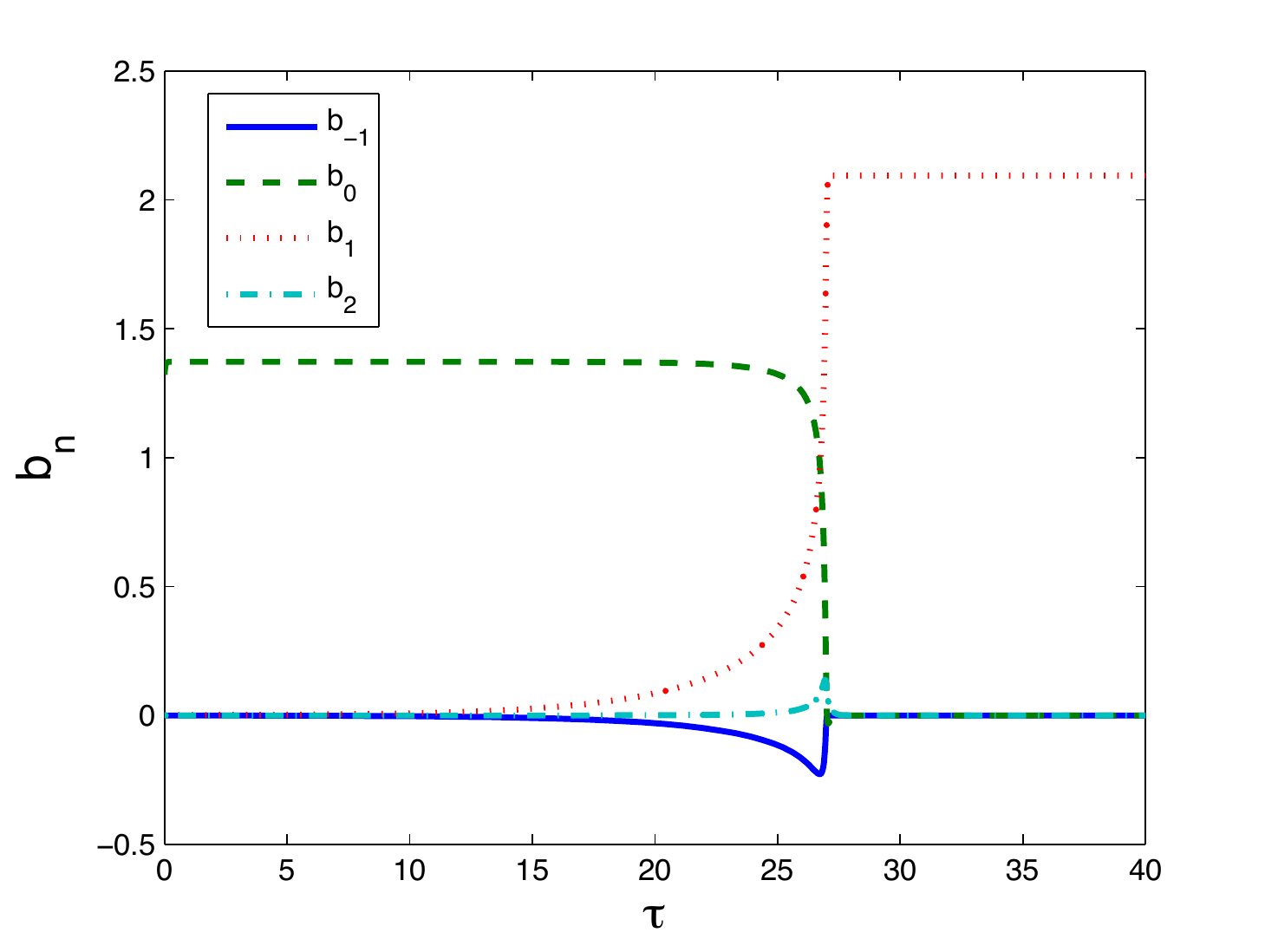}}%
\subfigure[Saddle-node bifurcation]{\includegraphics[width=0.5\textwidth]{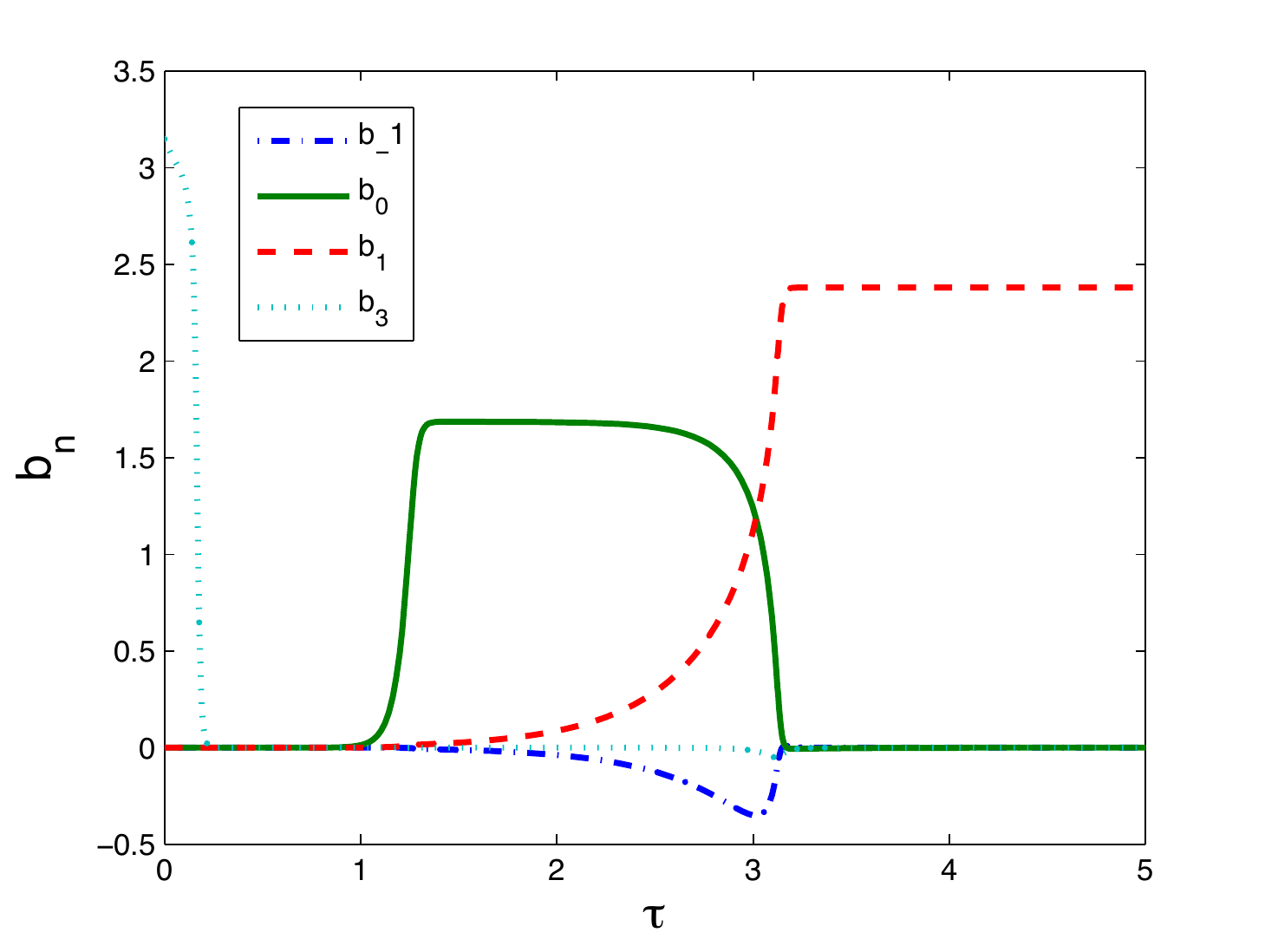}}%
\caption{\label{mode amplitudes} Time evolution of the amplitudes of
  the four largest modes that participate in (a) the transition from
  the initial $k_0$ pattern to the $k_1$ pattern, when the value of
  the control parameter is changed from $g=10$ to $g=11$, causing the
  initial $k_0$ pattern to experience an Eckhaus instability; and (b)
  the transition from the $k_{3}$ pattern to the $k_1$ pattern, when
  the value of the control parameter is changed from $g=20$ to $g=19$,
  causing the $k_3$ pattern to go through a saddle-node
  bifurcation. In both cases the results are obtained by a numerical
  integration of the seven truncated mode equations~(\ref{a-eqn}), for
  modes $b_{-3}$ to $b_{3}$, using the same parameters as in
  Fig.~\ref{StbBalloon}. Details of the transitions are discussed in
  the text. From Kenig \emph{et al.}  (2009$a$). Copyright (2009)
  American Physical Society.%
  \index{Eckhaus instability}\index{bifurcation!saddle-node}}
\end{figure}

We take a closer look at the transient behavior during the first
Eckhaus transition from the initial $k_0$ pattern to the $k_1$ pattern
by plotting the time evolution of the four largest modes, as shown in
Fig.~\ref{mode amplitudes}(a). One can observe the decay of the
unstable mode amplitude $b_{0}$ followed by the growth of $b_{1}$ to
its steady-state value. One can also see that during the transient the
amplitude of the unstable mode $b_{-1}$ becomes non-zero. Its
participation in the Eckhaus transition from the $k_0$ pattern to the
$k_1$ pattern is essential, as can be verified by considering these
two modes alone in a truncated expansion.  Limiting the expansion to
$b_0$ and $b_1$ suppresses the Eckhaus transition, and the $k_0$
pattern remains stable as $g$ exceeds its expected value for the
Eckhaus instability. The Eckhaus transition is observed only when the
$k_{-1}$ mode is included as well, corresponding to the stability
calculation, performed earlier for the state given by
Eq.~(\ref{BStability}).

One might naively expect that the same mechanism causes the transition
from the $k_{3}$ pattern to the $k_{1}$ pattern at $g=19$ through a
double phase slip (with $Q=2\Delta Q_N$), however, this is not the
case.  Fig.~\ref{mode amplitudes}(b) reveals the transient processes
on a downward sweep of $g$ just below the saddle node at $g=19$. As
$g$ crosses the\index{bifurcation!saddle-node} saddle-node bifurcation
point, the amplitude $b_{3}$ drops abruptly to zero. As can be seen
from Eq.~(\ref{a-eqn}), in the zero-displacement state the linear
growth rates of the solutions (\ref{multimode}) are $g-k_{n}^{2}$, so
the $k_{0}$ pattern has the largest possible growth rate and it
out-grows the other modes until its amplitude approaches the steady
state value (\ref{Bsteady}).  However, at this value of $g$ the
$k_{0}$ pattern is Eckhaus unstable with respect to the $k_{1}$
pattern---notice the characteristic evolution of the modes around
$\tau=3$ in Fig.~\ref{mode amplitudes}(b) corresponding to the Eckhaus
instability [cf.\ around $\tau=25$ in Fig.~\ref{mode amplitudes}(a)].
Thus the $k_{1}$ mode is ultimately the selected pattern.

For more rapid increases in the control parameter larger jumps in the
mode number may occur, and these were shown to be predicted simply
from a linear stability analysis following the Eckhaus
instability. \shortciteN{kenig09} showed that following an abrupt
increase of $g$ that crosses the Eckhaus instability line it is simply
the mode whose linear growth rate is greatest that is selected. For a
slow temporal ramp of the control parameter $g=\alpha\tau$, with
$\alpha\ll1$, they encountered a more interesting competition between
the different patterns. They showed that as the control parameter is
ramped a sequence of patterns start to grow one by one, yet the growth
rates increase with each pattern that emerges. This resembles a
balanced race in which the slow runners are allowed to start running
before the fast ones. Nevertheless, \shortciteN{kenig09} were able to
predict the winning pattern, and its dependence on the ramp rate
$\alpha$.  In all cases that were checked, simulations of the original
equations of motion of the resonators~(\ref{eq:bcleom}) confirm the
results based on the BCL amplitude equation.

\index{BCL equation|)}

\section{Continuous Amplitude Equations:\\
Example IV -- Intrinsic localized modes (ILMs)}
\label{sec:ILM}
\index{localized modes|(}
\index{nonlinear Schr\"{o}dinger equation!parametrically driven|(}
\index{PDNLS equation|(}

\subsection{Derivation of the PDNLS equation}

As our final example we focus on a different type of nonlinear states,
namely, intrinsic localized modes (ILMs), also known as discrete
breathers or lattice
solitons~\shortcite{Ovchinnikov82,sievers88,campbell04,Maniadis06}. These
localized states are intrinsic in the sense that they arise from the
inherent nonlinearity of the resonators, rather than from
extrinsically-imposed disorder as in the case of Anderson
localization. ILMs were observed by Sato \emph{et al.}~\citeyear{%
  sato03_1,sato03_2,sato04,sato06,sato07,sato08} in driven arrays of
micromechanical resonators. They were also observed in a wide range of
other physical systems including coupled arrays of Josephson
junctions~\shortcite{trias,binder00}, coupled optical
waveguides~\shortcite{eisenberg,eisenberg01,shimshon1},
two-dimensional nonlinear photonic crystals~\shortcite{fleischer},
highly-nonlinear atomic lattices~\shortcite{swanson}, and
antiferromagnets~\shortcite{schwarz,satoAFM04}. Thus, the ability to
perform a quantitative comparison between our theory and future
experiments with large arrays of MEMS and NEMS resonators, may have
consequences far beyond the framework of mechanical systems considered
here.

We follow the work of \shortciteN{kenigILM}, whose goal was to predict
the actual physical parameters, in realistic arrays of MEMS and NEMS
resonators, for which ILMs can form and sustain themselves. Such
predictions may have practical consequences for actual applications
exploiting self-localization to focus energy, and others that may want
to avoid energy focusing, for example in cases where very large
oscillation amplitudes may lead to mechanical failure. Again we wish
to formulate our analysis in terms of a continuous amplitude equation
and to display the range of stable ILMs on a reduced diagram---as we
did for extended modes in Fig.~\ref{StbBalloon}---helping to describe
the general qualitative behavior as physical parameters are varied.

We start with the same form of the equations of
motion~(\ref{eq:bcleom}) that we used in the preceding section, with
two differences: (1) We wish to keep an explicit parameter with which
we can vary the linear damping, thus we define the small expansion
parameter as $Q^{-1}=\epsilon\hat{\gamma}$, with $\epsilon\ll1$, and
$\hat\gamma$ of order unity; and (2) We use a negative sign before the
coupling coefficient $D$ to model\index{coupling!elastic} elastic
coupling between adjacent beams, which is stronger as the separation
between neighbors increases, thus acting to stiffen the
resonators. This leads to a dispersion\index{dispersion relation}
curve that has a positive slope, or a positive group velocity.%
\index{group velocity} The coupling mechanism in the experimental
setups in which ILMs were observed by Sato \emph{et
  al.}~\citeyear{sato03_1,sato03_2,sato04,sato06,sato07,sato08} is of
this kind. Our equations of motion then become
\index{equation of motion!one-dimensional array!scaled}
\begin{equation}
  \label{eq:ilmeom}
  \ddot{u}_{n} + \epsilon\hat{\gamma}\dot{u}_{n} + \left(1 - \epsilon
  \hat{h}\cos2\omega_{p}t\right)u_{n} 
  - \tfrac{1}{2} D(u_{n+1}-2u_{n}+u_{n-1}) + u_{n}^{3} 
  + \hat{\eta} u_{n}^{2}\dot{u}_{n}=0,
\end{equation}
with hats to be removed later by additional scaling.

An experimental protocol for producing ILMs in an array of resonators
with a stiffening nonlinearity---albeit not the one we use below---is
to drive the array at the highest-frequency extended mode.  As the
resonators are collectively oscillating at this mode, the frequency is
raised further, which through the stiffening Duffing
nonlinearity\index{Duffing nonlinearity} results in an increase of the
oscillation amplitude\index{frequency pulling} up to a point at which
the extended pattern breaks into localized
modes~\shortcite{sato03_1,sato06}.  With this in mind---and
concentrating on the case of elastic coupling where the
highest-frequency mode $\omega = \sqrt{1+2D}$ is the staggered mode,
in which adjacent resonators oscillate out of phase---we write the
displacement of the $n^{th}$ resonator as%
\index{multiple scales!ansatz}
\begin{equation}
  \label{u_expansion}
    u_{n} = \epsilon^{1/2} \bigl[\hat{\psi}(\hat{X}_n,\hat{T})
    e^{i(\omega t - \pi n)} + c.c.\bigr]
    + \epsilon^{3/2}u_{n}^{(1)}(t,\hat{T},\hat{X}_n) + \ldots,
\end{equation}
with slow temporal and spatial variables $\hat{T}=\epsilon t$ and
$\hat{X}_n = \epsilon^{1/2} n$. As usual, we take the parametric drive
frequency to be close to twice $\omega$ by setting $\omega_{p} =
\omega +\epsilon \Omega/2$, introduce a continuous spatial variable
$\hat{X}$ in place of $\hat{X}_n$, and substitute the ansatz
(\ref{u_expansion}) into the equations of motion (\ref{eq:ilmeom}) term by
term. Up to order $\epsilon^{3/2}$ we have
\begin{subequations}
\begin{align}
  &\ddot{u}_{n} = \epsilon^{1/2} \left[\left(-\omega^{2}\hat{\psi} +
    2i\omega\epsilon\frac{\partial\hat{\psi}}{\partial\hat{T}}\right)
  e^{i(\omega t - \pi n)}  + c.c.\right] +
\epsilon^{3/2}\ddot{u}^{(1)}_{n},\\ 
  &u_{n\pm1} = -\epsilon^{1/2} \Bigg[\left(\hat{\psi} \pm \epsilon^{1/2}
    \frac{\partial\hat{\psi}}{\partial\hat{X}} + \frac{\epsilon}{2}
    \frac{\partial^{2}\hat{\psi}}{\partial\hat{X}^{2}}\right)
  e^{i(\omega t-\pi n)} + c.c.\Bigg] + \epsilon^{3/2}u^{(1)}_{n\pm1},\\
  &\epsilon \hat{h}\cos(2\omega_{p}t)u_{n} =
  \epsilon^{3/2}\frac{\hat{h}}{2}\hat{\psi}^{*}e^{i\Omega\hat{T}}e^{i(\omega
  t+\pi n)}  + O(e^{i3\omega t}) + c.c., \\
  &\epsilon\hat{\gamma}\dot{u}_{n} =
  \epsilon^{3/2}\hat{\gamma}i\omega\hat{\psi}e^{i(\omega t - \pi n)} +
  c.c.,  \\
  &u_{n}^{3} =  \epsilon^{3/2}3|\hat{\psi}|^{2}\hat{\psi}e^{i(\omega
  t - \pi n)}  + O(e^{i3\omega t},e^{i3\pi n}) + c.c.,
\end{align}
and
\begin{align}
  &u_{n}^{2}\dot{u}_{n} =
  \epsilon^{3/2}i\omega|\hat{\psi}|^{2}\hat{\psi}e^{i(\omega t-\pi n)}
   + O(e^{i3\omega t},e^{i3\pi n}) + c.c.,\hphantom{xxxxxxxxxxxxx}
\end{align}
\end{subequations}
where $O(e^{i3\omega t},e^{i3\pi n})$ are fast oscillating terms with
temporal frequency $3\omega$ or spatial wavenumber $3\pi$.

At order $\epsilon^{1/2}$ the equations of motion~(\ref{eq:ilmeom})
are satisfied trivially. However, once again at order $\epsilon^{3/2}$
we encounter secular terms\index{secular term}---in this case,
proportional to $e^{i(\omega t-\pi n)}$--- and must apply a
\emph{solvability condition}, \index{solvability condition} requiring
all such terms to vanish. Again, it is this condition that leads to a
partial differential equation (PDE) describing the slow dynamics of
the amplitudes of the resonators,
\begin{equation}
  \label{sol_cond}
  2i\omega\frac{\partial \hat{\psi}}{\partial \hat{T}} + (3 +
  i\omega\hat{\eta})|\hat{\psi}|^{2}\hat{\psi} +
  \frac{1}{2}D\frac{\partial^{2}\hat{\psi}}{\partial \hat{X}^{2}} +
  i\hat{\gamma}\omega\hat{\psi} -
  \frac{\hat{h}}{2}\hat{\psi}^{*}e^{i\Omega \hat{T}} = 0.
\end{equation}
Note that while $e^{i(\omega t+\pi n)}=e^{i(\omega t-\pi n)}$, if we
were to consider an arbitrary mode of wave number $q$ instead of
$\pi$, the parametric term would have forced us to apply another
solvability condition, requiring terms proportional to $e^{i(\omega
  t+qn)}$ to vanish. This was exactly the situation in the preceding
section \ref{sec:patterns}, where we were forced first to consider an
ansatz based on counter propagating waves as the $O(\epsilon^{1/2})$
solution for $u_{n}$, leading to a system of two coupled amplitude
equations~(\ref{AmpEqs}), after which a second scaling was used to
obtain a single amplitude equation~(\ref{BampEq}). Here we can get
away with a single step.

By means of rescaling,
\begin{gather}
  \hat{\psi} = \sqrt\frac{2 \omega\Omega}{3}\psi,\quad
  \hat{X}=\sqrt\frac{D}{2\omega\Omega} X,\quad
  \hat{T}=\frac{2}{\Omega} T,\quad
  \hat{h}=2\omega\Omega h,\quad
  \hat{\gamma}=\Omega \gamma,\quad
  \hat{\eta}=\frac{3}{2\omega} \eta,
  \label{scaling}
\end{gather}
we transform \eqref{sol_cond} into a normalized form,
\begin{equation}
  \label{PDNLS}
  i\frac{\partial \psi}{\partial T} =
  - \frac{\partial^{2} \psi}{\partial X^{2}}
  - i\gamma\psi - (2 + i\eta)|\psi|^{2}\psi + h\psi^{*}e^{2iT}.
\end{equation}
We then perform one final transformation $\psi\rightarrow\psi e^{iT}$ and
arrive at an autonomous PDE, which is the final form of our amplitude equation,
\index{amplitude equation!continuous}
\index{nonlinear Schr\"{o}dinger equation!parametrically driven!with nonlinear damping}%
\index{nonlinear Schr\"{o}dinger equation!parametrically driven|textbf}
\index{PDNLS equation|textbf}
\begin{equation}
  \label{amp_eq}
  \boxed{
    i\frac{\partial \psi}{\partial T} =
    - \frac{\partial^{2} \psi}{\partial X^{2}}
    + (1 - i\gamma)\psi - (2 + i\eta)|\psi|^{2}\psi + h\psi^{*}
  }\ .
\end{equation}

Equation (\ref{PDNLS}) with $\eta=0$ is called the parametrically
driven damped nonlinear Schr\"odinger equation (PDNLS).  It models
parametrically driven media in
hydrodynamics~\shortcite{zhang,wang97,wang98,miao} and
optics~\shortcite{longhi96,sanchez}, and was also used as an amplitude
equation to study localized structures in arrays of coupled
pendulums~\shortcite{denardo,chen,barashenkov00}. Recently, a pair of
linearly-coupled PDNLS equations was used to model coupled dual-core
wave guides~\shortcite{dror}. Equation~(\ref{amp_eq}) has the form of a
\index{Ginzburg-Landau equation|textbf} forced complex Ginzburg-Landau
equation~\shortcite{yochelis} but with specific coefficients that are
derived, via the scaling performed in~(\ref{u_expansion}) and
(\ref{scaling}), from the underlying equations of
motion~(\ref{eq:ilmeom}).
We note that considering the equations of motion (\ref{LCeom}) (yet
still with a negative sign before $D$) as our stating point instead of
eqns~(\ref{eq:ilmeom}) leads to the same \eqref{sol_cond} as above,
but with slightly different coefficients. Thus, applying modified
scaling~(\ref{scaling}) yields exactly the same amplitude
equation~(\ref{amp_eq}).

\subsection{Analyzing and solving the equation}

\index{soliton solution|(}
\index{nonlinear Schr\"{o}dinger equation!parametrically driven!with linear damping}%

\begin{figure}[b]
\begin{center}
\includegraphics[width=0.6\columnwidth]{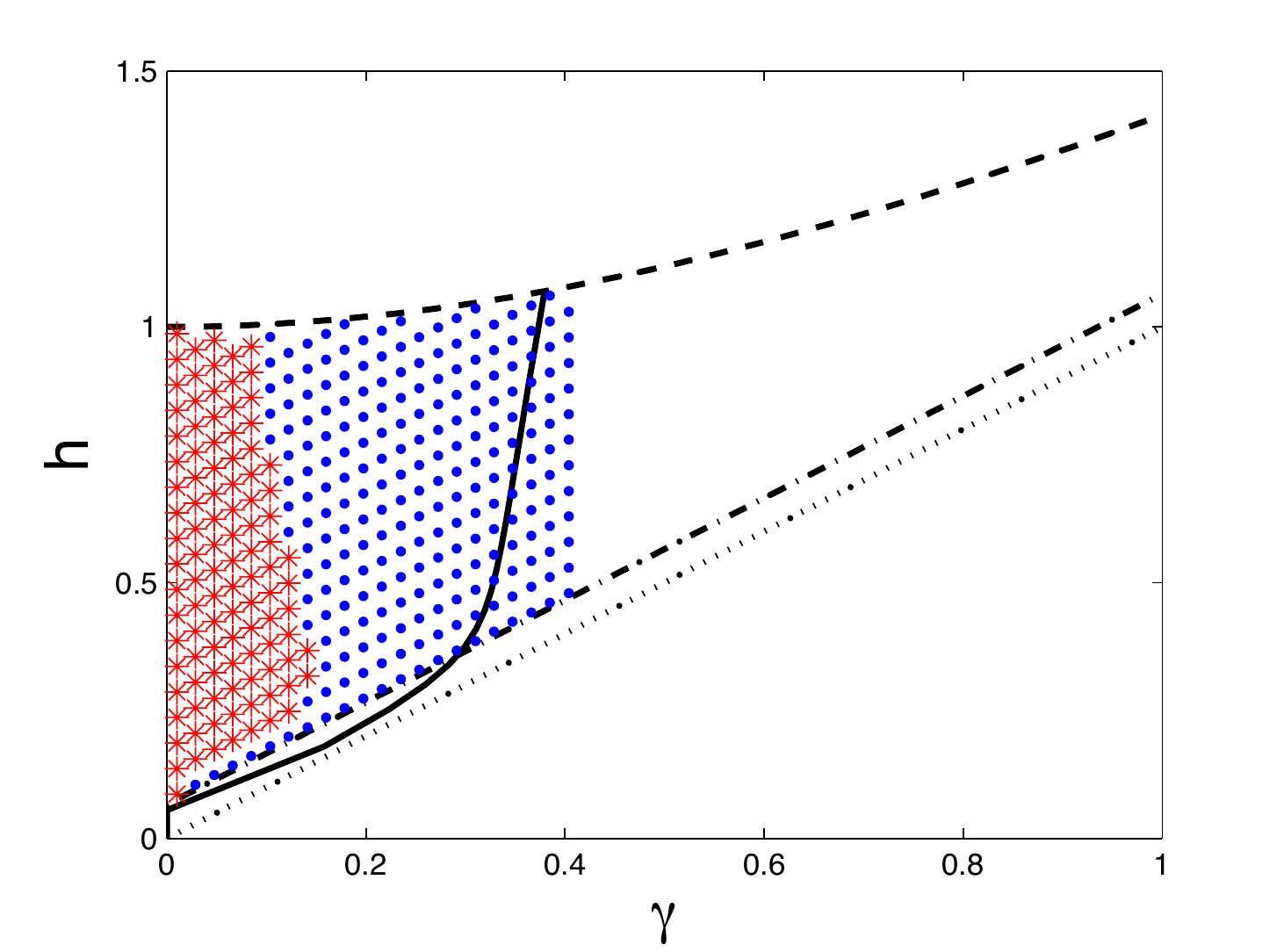}%
\caption{Stability diagram for localized solutions of
  the amplitude equation~(\ref{amp_eq}) in the $h$ vs.~$\gamma$ plane.
  The dotted line is the lower existence boundary for $\eta=0$, namely
  $h=\gamma$. The dash-dotted line is the approximate low boundary
  for $\eta=0.1$, given by \eqref{sn_g}. Above the solid line the
  $\Psi_{+}$ solution of the PDNLS equation with $\eta=0$ is unstable
  with respect to a Hopf bifurcation. 
  The dashed line is the line $h=\sqrt{1+\gamma^{2}}$ above which the zero
  solution is unstable. Red $\ast$s are points for which linear
  stability analysis shows that perturbations away from the soliton
  solution $\psi(X)$ grow exponentially for $\eta=0.1$, hence the soliton solution
  is unstable. Blue dots represent points for which the
  solution $\psi(X)$ is stable according to the linear analysis. From
  Kenig \emph{et al.} (2009$b$). Copyright (2009) American Physical Society.
  \index{stability diagram}
  \label{stability_diagram}}
\end{center}
\end{figure}

A remarkable feature of the amplitude equation (\ref{amp_eq}) is that
for $\eta=0$ it has exact steady-state%
\index{steady-state solution!scaled} solitonic solutions, as shown by
\shortciteN{barashenkov91},\index{soliton solution}
\begin{equation}
  \label{NLS_sol}
  \Psi_{\pm}(X) = A_{\pm}e^{-i\Theta{\pm}}
  \textrm{sech}\left[A_{\pm}\left(X - X_{0}\right)\right],
\end{equation}
where $X_{0}$ is an arbitrary position of the soliton, 
\begin{equation}
  \label{NLS_amp}
  A^{2}_{\pm} = 1\pm\sqrt{h^2 - \gamma^2},{\rm \ and \ }
  \cos(2\Theta_{\pm}) =\pm\sqrt{1 - \frac{\gamma^{2}}{h^{2}}}.
\end{equation}
This pair of solitonic solutions exists for $\gamma<h$, above the
dotted line in Fig.~\ref{stability_diagram}.  It was shown by
\shortciteN{barashenkov91} that the $\Psi_{-}$ soliton is unstable for
all values of $\gamma$ and $h$, while the $\Psi_{+}$ soliton is stable
in a certain parameter range.  A simple linear stability analysis
shows that the zero solution $\psi(X)=0$, which exists for all
parameter values, is stable only for $h<\sqrt{1+\gamma^{2}}$. This
inequality, indicated by a dashed line in
Fig.~\ref{stability_diagram}, also determines an upper stability limit
for localized solutions of \eqref{amp_eq}, as they decay exponentially
to zero on either side.

We follow \shortciteN{kenigILM} in constructing an approximate
analytical expression for the localized solution of the full amplitude
equation (\ref{amp_eq}), with $\eta>0$, implementing a method
introduced by \shortciteN{barashenkov03}. To this end, we consider a
function of the same form as $\Psi_{\pm}$,
\begin{equation}
  \label{aprx_sol_time_dependant}
  \psi(X,T) = a(T) e^{-i\theta(T)}
  \textrm{sech}\left[a(T) \left(X-X_0\right)\right],
\end{equation}
except that $a$ and $\theta$ are now time-dependent. We multiply
Eq.~(\ref{amp_eq}) by $\psi^{*}$, subtract the complex conjugate of
the resulting equation and get
\begin{equation}
  i\frac{\partial|\psi|^{2}}{\partial T} = -\frac{\partial}{\partial
    X}{\left(\frac{\partial \psi}{\partial X}\psi^{*} -
      \psi\frac{\partial \psi^{*}}{\partial X}\right)} +
  h[(\psi^{*})^{2} - \psi^{2}]
  - 2i\gamma|\psi|^{2} - 2i\eta|\psi|^{4}.
\end{equation}
By substituting $\psi=|\psi|e^{-i\chi}$, integrating over $X'=X-X_0$,
and assuming that $\psi\rightarrow0$ and $\partial\psi/\partial
X\rightarrow0$ as $|X|\rightarrow\infty$, we obtain a
spatially-independent integral equation
\begin{equation}
  \label{no_x_eq}
  \frac{d}{dT}\int|\psi|^{2}dX' = 2\int|\psi|^{2}[h\sin(2\chi) -
  \gamma]dX' - 2\eta\int|\psi|^{4}dX'.
\end{equation}
Substituting the ansatz~(\ref{aprx_sol_time_dependant}) into
\eqref{no_x_eq}, we obtain the time evolution equation for $a$
\begin{equation}
  \label{a_T}
  \frac{d a}{dT}=2a(h\sin(2\theta) - \gamma - \tilde{\eta}a^{2}),
\end{equation}
where $\tilde{\eta}=2\eta/3$. The time evolution equation for $\theta$
is derived in a similar way by multiplying \eqref{amp_eq} by
$\psi^{*}$, adding the complex conjugate of the resulting equation,
substituting the ansatz~(\ref{aprx_sol_time_dependant}), and
integrating over space to yield
\begin{equation}
  \label{theta_T}
  \frac{d\theta}{dT} = h\cos(2\theta) + 1 - a^{2}.
\end{equation}

Equations (\ref{a_T}) and (\ref{theta_T}) have the same form as the
equations obtained by \shortciteN{barashenkov03}, whose fixed points
are
\begin{equation}
  \label{aprx_amp}
  a^{2}_{\pm} = \frac{1 - \gamma\tilde{\eta}
    \pm\sqrt{h^{2}(1 + \tilde{\eta}^{2}) - 
      (\gamma + \tilde\eta)^2}}{1 + \tilde{\eta}^{2}},
\end{equation}
which has to be positive, and
\begin{eqnarray}
  \label{aprx_phase}\nonumber
  h\cos(2\theta_{\pm}) &= &a_{\pm}^{2} - 1,\\
  h\sin(2\theta_{\pm}) &= &\gamma + \tilde\eta a_{\pm}^{2}.
\end{eqnarray}
A linear analysis of these stationary points shows that
$(a_{+},\theta_{+})$ and $(a_{-},\theta_{-})$ are a stable node and a
saddle, respectively~\shortcite{barashenkov03}. The saddle-node
bifurcation\index{bifurcation!saddle-node} point of these solutions
occurs at
\begin{equation}
  \label{sn_g}
  h_{sn}(\tilde\eta) = \frac{\gamma + \tilde{\eta}}{\sqrt{1 +
  \tilde{\eta}^{2}}}, 
  \quad \textrm{where} \quad \tilde{\eta}=\frac{2}{3} \eta,
\end{equation}
as long as $\gamma\tilde\eta<1$.  This is the approximate minimal
driving strength required to support a localized structure in the
array, in the presence of linear and nonlinear dissipation. It is
indicated by a dash-dotted line in Fig~\ref{stability_diagram}, for
$\eta=0.1$.

The approximate stable localized solution of the amplitude
equation~(\ref{amp_eq}) is therefore given by
\begin{equation}
  \label{aprx_sol}
  \pap(X) = a_{+}e^{-i\theta_{+}}\textrm{sech}(a_{+}(X - X_{0})).
\end{equation}
Substituting this expression into \eqref{u_expansion} yields an
approximate expression for the displacements of the actual resonators
in the array of the form\index{steady-state solution}%
\index{soliton solution!approximate with nonlinear damping}  
\begin{equation}
  \label{u_sol}
  u_{n}(t) \simeq 2\sqrt{\frac{2\epsilon\omega\Omega}{3}}a_{+}
  \textrm{sech}\left[a_{+}\left(\sqrt{\frac{2\epsilon\omega\Omega}{D}}n
      - X_{0}\right)\right] \cos\left(\omega_{p}t - \pi n - \theta_{+}\right).
\end{equation}

To obtain accurate solutions one has no choice but to solve the
amplitude equation, or the underlying discrete equations of motion,
numerically. We do so with the equations of motion~(\ref{eq:ilmeom})
by initiating them with the approximate expression~(\ref{u_sol}) at a
value of $h$ just above the saddle node $h_{sn}$~(\ref{sn_g}). We then
perform a quasistatic upward sweep%
\index{quasistatic sweep!of drive amplitude} of $h$, raising $h$ in
small increments and waiting for transients to decay at each step.  To
obtain the stationary solution of the amplitude equation we set
$\partial\psi/\partial T=0$ in \eqref{amp_eq} and solve it numerically
as a boundary value problem over an interval of length $L$, with
boundary conditions $\psi(X=0)=\psi(X=L)=0$.%
\index{boundary conditions!fixed} We use the approximate expression
$\pap(X)$ [\eqref{aprx_sol}] as an initial guess.  Having identified
an upper stability boundary $h=\sqrt{1+\gamma^{2}}$ and an approximate
lower existence boundary, given by Eq.~(\ref{sn_g}), we must make use
of the numerically obtained localized solutions $\psi(X)$ in order to
examine the stability within these boundaries [see
\shortciteN{kenigILM} for details]. The stability diagram of both the
analytical solution $\Psi_{+}$ for $\eta=0$~\shortcite{barashenkov91}
and the numerical solution $\psi(X)$ for $\eta=0.1$ are displayed in
Fig.~\ref{stability_diagram}. Numerical integration of the underlying equations of
motion~(\ref{eq:ilmeom}) confirms the stability analysis, based on the
amplitude equation \shortcite{kenigILM}.

\index{damping!nonlinear|(}
Figure~\ref{stability_diagram} highlights the effects of nonlinear
damping on localized solutions. The first effect is to raise the lower
existence boundary. This is explained by the fact that the additional
energy lost through nonlinear damping has to be compensated by an
increase in the strength of the parametric drive, as predicted by the
approximate expression~(\ref{sn_g}).  The second effect is that
nonlinear damping increases the area in the $(h,\gamma)$ parameter
space where solitons are stable (blue dots). In particular, the shape
of the unstable region for $\eta>0$ (red $\ast$s) becomes
qualitatively different. There are values of $\gamma$ for which an
increase in the drive amplitude $h$ initially induces an instability
of the soliton, while upon further increase of $h$ the soliton regains
its stability. This can be explained by noting that the amplitude of
the soliton---given approximately by Eq.~(\ref{aprx_amp})---increases
as $h$ becomes larger, thereby enhancing the effect of nonlinear
damping. This increase of damping exerts a similar stabilizing effect
as that of increasing $\gamma$ in the absence of nonlinear damping.

\index{damping!nonlinear|)}

\subsection{Brief survey of applications}

\begin{figure}[b]
    \centering
    \subfigure{
    \includegraphics[width=0.45\columnwidth]{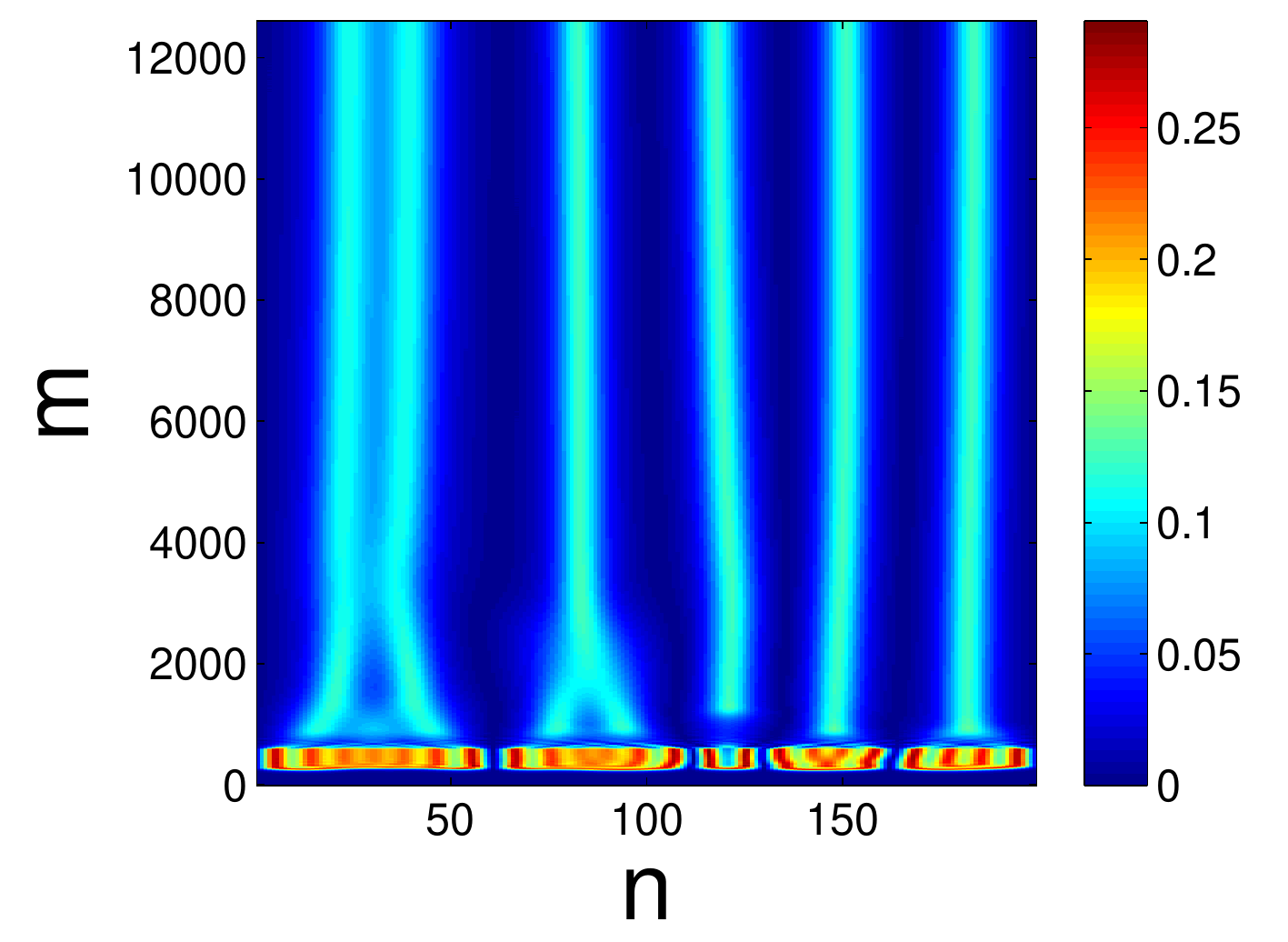}}
    \subfigure{
    \includegraphics[width=0.45\columnwidth]{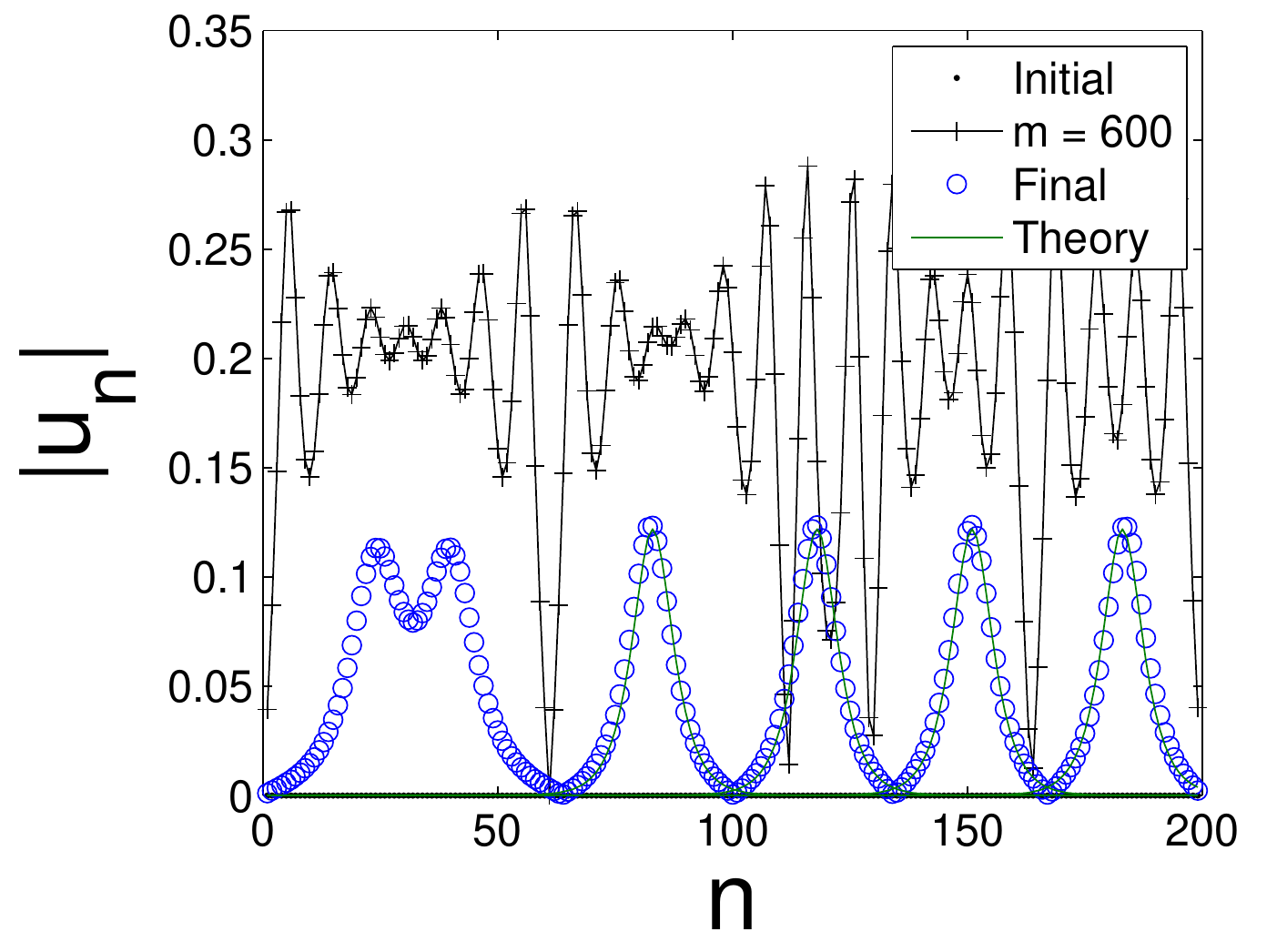}}
    \subfigure{
    \includegraphics[width=0.45\columnwidth]{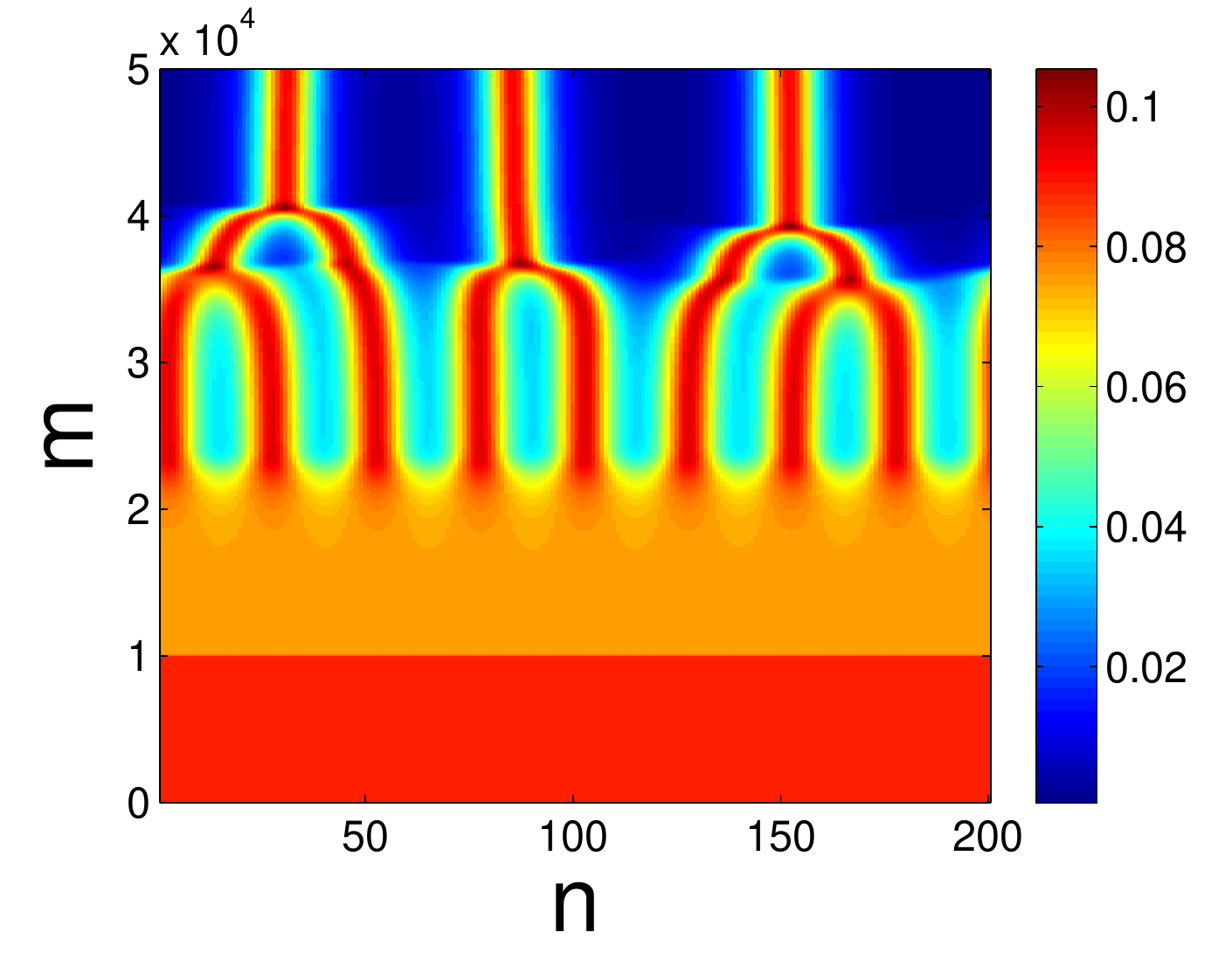}}
    \subfigure{
    \includegraphics[width=0.45\columnwidth]{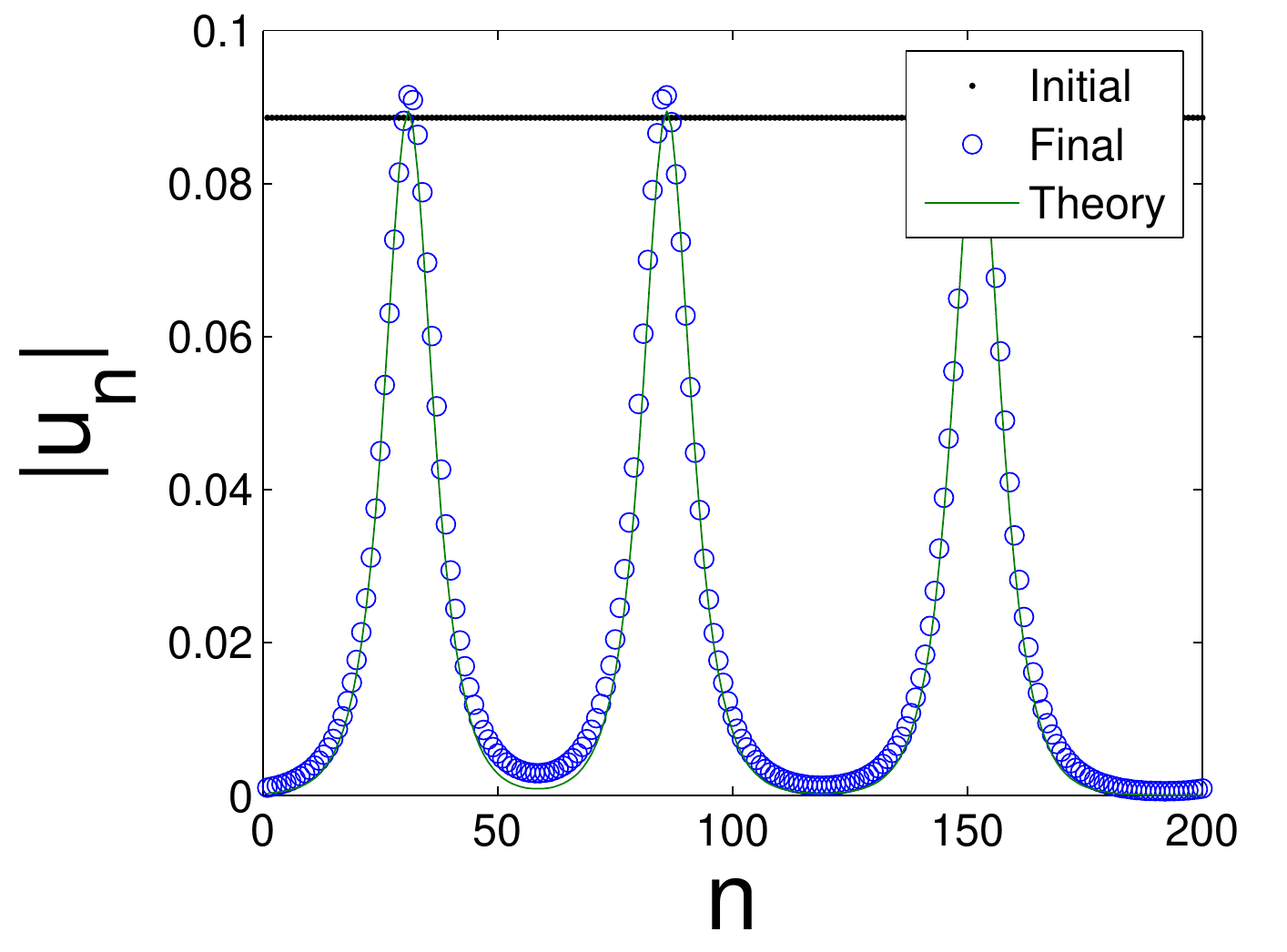}}
  \caption{\label{ilm creation}(Color) Numerical simulation of the
    equations of motion~(\ref{eq:ilmeom}) showing the dynamical
    creation of solitons, for $\gamma=1$, $\eta=0.3$, $D=0.25$,
    $\epsilon=0.01$, and $\omega_p=1.002\omega$.  Plotted are the
    absolute values of the displacements of the resonators, which
    alternate between positive and negative values.  Left panels show
    the time evolution, with $m$ counting the number of drive
    periods. Right panels show the initial (black dots) and final
    (blue circles) states along with the analytical form of the
    solitons (green solid line), using only their central positions
    $X_0$ as fitting parameters. \emph{Top panels}: A simulation of
    199 resonators with fixed boundary conditions is initiated with
    random noise and a drive amplitude of $h=5$, which is above the
    upper stability limit, $h=\sqrt{1+\gamma^{2}}=\sqrt2$, for both
    the zero-state and the solitons.  At time $m=600$ drive periods,
    after a non-zero transient (black $+$s in the right panel) has
    developed, the drive amplitude is lowered to $h=1.35<\sqrt2$,
    yielding stable solitons.  \emph{Bottom panels}: A simulation of
    200 resonators with periodic boundary conditions\index{boundary
      conditions!periodic} is initiated with the uniform non-zero
    solution and a drive amplitude of $h=1.3$, which is above the
    calculated stability threshold, $h_{th}\simeq1.26$, for this state
    (Kenig \emph{et al.} 2009$b$).  After $m=10000$ drive periods
    during which the uniform state remains stable, the drive amplitude
    is lowered to $h=1.2<h_{th}$, yielding stable solitons. From Kenig
    \emph{et al.} (2009$b$). Copyright (2009) American Physical Society.}
\end{figure}

\shortciteN{kenigILM} studies a host of dynamical phenomena using
their version of the PDNLS equation~(\ref{amp_eq}). These include
questions concerning the interaction of pairs of solitons, which can
be either attractive or repulsive, depending on the relative phase of
the two solitons; the possibility of pairs of solitons to form bound
states; and the ability of a boosted soliton spontaneously to split
into two.  We urge the reader to consult the original work for
additional detail, and only emphasize the exquisite agreement between
the predictions made with the amplitude equation~(\ref{amp_eq}), and
numerical simulations of the underlying equations of
motion~(\ref{eq:ilmeom}).

There is one particular question that we wish to address here which
deals with the procedure for generating solitons in actual
experiments, where one cannot simply introduce an approximate soliton
as an initial condition like one does in a simulation. Moreover,
it is not obvious how dynamically to form solitons starting with a
motionless array of resonators, as one needs to take the system
sufficiently far from the basin of attraction of the zero solution
$\psi(X)=0$, which is also stable whenever solitons are stable. The
most direct procedure for avoiding the zero solution, starting from
weak random noise, is to drive the system with
$h>\sqrt{1+\gamma^{2}}$, so neither the zero solution nor the soliton
solutions are stable. As a consequence, a non-zero pattern develops.
Stable solitons can then be formed by lowering the drive amplitude to
a value $h<\sqrt{1+\gamma^{2}}$ for which the zero solution and the
soliton solutions are both stable, if the non-zero pattern that was
obtained is outside the basin of attraction of the zero solution.

This simple procedure, sometimes called \emph{self trapping}---which%
\index{soliton solution!self trapping} could be implemented
experimentally in a straightforward manner---is demonstrated in the
top panels of Fig.~\ref{ilm creation}, showing a numerical simulation
of the equations of motion~(\ref{eq:ilmeom}) with fixed boundary
conditions, using $N=199$ resonators. One can see that the initial
transient that forms becomes unstable upon lowering the drive
amplitude, giving rise to the formation of a number of solitons. Note
that before reaching steady state, a pair of solitons merges into one,
and another pair attracts and forms a bound state. Both of these
effects were studied by \shortciteN{kenigILM}. The emerging isolated
solitons agree well with the approximate analytical
form~(\ref{u_sol}), determined earlier, with only their central
positions $X_0$ used as fitting parameters.

A more controlled procedure for generating solitons would be to
initiate the array in a particular non-zero state and then to drive it
outside its known stability boundaries. This has been considered in
the past in systems without nonlinear damping, using the non-zero
uniform solution of the PDNLS~\shortcite{barashenkov03}.  However, it
is known for systems with $\eta=0$ that the uniform solution is always
unstable against weak modulations and so may be difficult to access
dynamically. What \shortciteN{kenigILM} discovered was that nonlinear
damping can act to stabilize the non-zero spatially-uniform solution,
making the procedure for generating solitons through a modulation
instability of a uniform state possibly relevant for experiments.%
\index{soliton solution!modulation instability of uniform state}

We demonstrate the use of the stable uniform solution in the dynamical
formation of solitons in the bottom panels of Fig.~\ref{ilm creation},
showing a numerical simulation of the equations of
motion~(\ref{eq:ilmeom}) with\index{boundary conditions!periodic}
periodic boundary conditions, using $N=200$ resonators. The array is
initiated with the large-amplitude uniform solution and is driven
within the stability boundary of this state, as calculated from the
amplitude equation~(\ref{amp_eq}). After a long time during which the
uniform solution indeed remains stable, the drive amplitude is lowered
below the stability threshold for this solution, but within the
stability boundaries of the soliton solutions, and solitons are formed
via a modulation of the unstable uniform state. 

\index{soliton solution|)}
\index{localized modes|)} 
\index{nonlinear Schr\"{o}dinger equation!parametrically driven|)}
\index{PDNLS equation|)}

\section*{Acknowledgments}

This work was supported by the US-Israel Binational Science Foundation
(BSF) through Grant No.~2004339, by the NSF through Grant
No.~DMR-1003337, and by the German-Israeli Foundation (GIF) through Grant
No.~981-185.14/2007.

\bibliographystyle{OUPnamed_notitle}
\bibliography{arrays}
\printindex

\end{document}